\documentclass[aps,prl,reprint,groupedaddress,twocolumn,superscriptaddress]{revtex4-1}
\usepackage{graphicx}     
\usepackage{subfigure}
\usepackage{dcolumn}     
\usepackage{bm}              
\usepackage{hyperref}     
\usepackage{amsmath}   
\usepackage{amssymb}
\usepackage{bbold}
\usepackage{pifont}
\usepackage{tabularx}
\usepackage{wasysym}
\usepackage{color}
\usepackage[table]{xcolor}
\usepackage{array}
\usepackage{multirow}
\usepackage{makecell}
\usepackage[version=3]{mhchem} 
\usepackage{float}
\usepackage{lipsum}
\usepackage{ulem}

\usepackage{verbatim}    
 
\makeatletter
\newcommand{\colorcaption}[2][]{%
  \begingroup%
  \renewcommand{\@caption@fignum@sep}{ (color online). }%
  \caption[#1]{#2}%
  \endgroup%
  }
\makeatother

\makeatletter
\newcommand*{\rom}[1]{\expandafter\@slowromancap\romannumeral #1@}
\makeatother

\begin{document}
\title{Topological Chiral Superconductivity Mediated by Intervalley Antiferromagnetic Fluctuations in Twisted Bilayer WSe$_2$}
\author{Wei Qin}
\email{qinwei5@ustc.edu.cn} 
\affiliation{Department of Physics, University of Science and Technology of China, Hefei, Anhui 230026, China}
\author{Wen-Xuan Qiu}
\thanks{These authors contributed equally to this work.}
\affiliation{School of Physics and Technology, Wuhan University, Wuhan 430072, China}
\author{Fengcheng Wu}
\email{wufcheng@whu.edu.cn}
\affiliation{School of Physics and Technology, Wuhan University, Wuhan 430072, China}
\date{\today}

\begin{abstract}
Motivated by the recent observations of superconductivity in twisted bilayer WSe$_2$ (tWSe$_2$), we theoretically investigate the superconductivity driven by electronic mechanism. We first demonstrate that the multi-band screened Coulomb interaction within the random phase approximation is insufficient to induce observable pairing instability. 
Nevertheless, by further including the intervalley antiferromagnetic fluctuations, the pairing interaction is substantially enhanced, yielding superconductivity with critical temperature $T_c$ of hundreds of millikelvin at van Hove singularities. The predicted $T_c$ increases with increasing the displacement field and corresponds to a doubly-degenerate $d$-wave-like pairing, which evolves into topological chiral $d \pm id$ superconductor below $T_c$. The interplay between superconductivity and intervalley antiferromagnetism results in a phase diagram consistent with
experimental observations.
These findings establish intervalley fluctuations as the primary pairing glue in tWSe$_2$.
\end{abstract}

\maketitle

\textit{Introduction.---}The discovery of superconductivity and correlated insulating states in magic-angle twisted bilayer graphene \cite{Cao:2018aa,Cao:2018ab} has stimulated extensive studies on the mori\'e superlattices \cite{Yankowitz:2019aa,Lu:2019aa,Balents:2020aa,Stepanov:2020aa,Saito:2020aa,Wong:2020aa,Zondiner:2020aa,Liu:2021aa,Park:2021aa,Hao:2021aa,Stepanov:2021aa,Saito:2021aa,Cao:2021aa,Choi:2021aa,Choi:2021ab,Jaoui:2022aa}, resulting in the observation of a plethora of intriguing strongly correlated phenomena. Beyond graphene-based systems, moir\'e transition metal dichalcogenides (TMDs) have emerged as another prominent class of moiré materials \cite{Wu:2018aa,Naik:2018aa,Wu:2019aa,Ruiz-Tijerina:2019aa,Schrade:2019aa}.  
Monolayer TMDs are semiconductors with strong spin-valley locking \cite{Xiao:2012aa}, and moir\'e  TMD bilayers can produce flat bands with enhanced interactions  \cite{Wu:2018aa,Naik:2018aa,Wu:2019aa}, leading to Mott insulators \cite{Regan:2020aa,Tang:2020aa,Wang:2020aa}, magnetism \cite{Xu:2022aa,Anderson:2023aa,Ghiotto:2024aa}, generalized Wigner crystals \cite{Xu:2020aa,Regan:2020aa,Li:2021ab}, and heavy fermion phenomena \cite{Zhao:2023aa}. The interplay of correlations and  topology has further enabled the discovery of integer and fractional Chern insulators \cite{Li:2021aa,Belanger:2022aa,Crepel:2023aa,Cai:2023aa,Zeng:2023aa,Park:2023aa,Xu:2023aa,Foutty:2024aa}, and the fractional quantum spin Hall insulator \cite{Kang:2024aa}. While both platforms host diverse correlated phases, robust superconductivity has been primarily established in the graphene-based systems.

Recently, robust superconductivity has been reported in tWSe$_2$ at two distinct twist angles: $\theta=3.65^{\circ}$ \cite{Xia:2024aa} and $\theta=5^{\circ}$ \cite{Guo:2024aa}. In the former case, superconductivity is observed at filling fraction $\nu=-1$ (one hole per moir\'e unit cell) under minimal displacement fields, with an optimal critical temperature $T_c$ of 0.22 K. The superconducting phase is in proximity to a correlated insulator phase at the same $\nu$ but finite displacement fields. By contrast, no prominent correlated insulator is observed at $\theta=5^{\circ}$, where the superconducting state emerges around the van Hove singularity with an optimal $T_c$ of 0.4 K under large displacement fields. In addition, the superconducting phase is adjacent to a metallic phase with enhanced resistivity attributed to Fermi surface (FS) reconstruction \cite{Guo:2024aa}. Although several models have been examined in relation to superconductivity in this system \cite{Hsu:2021aa,Zegrodnik:2023aa,Wu:2023aa,Schrade:2021aa,Akbar:2024aa,Christos:2024aa,Weber:2024aa,Tuo:2024aa,Kim:2024aa,Zhu:2024aa},
the pairing mechanism at a microscopic and quantitative level remains elusive. 

In this Letter, we investigate the mechanism of superconductivity driven solely by Coulomb interaction in tWSe$_2$ at $\theta=5^{\circ}$. Using the continuum band structure model with realistic material parameters \cite{Wu:2019aa,Devakul:2021aa}, we find that the Coulomb interaction is strongly screened within the multi-band random phase approximation (RPA), rendering it insufficient to induce detectable superconductivity in tWSe$_2$. However, by including intervalley antiferromagnetic (IVA) fluctuations on top of the RPA-screened Coulomb interaction, the pairing strength is significantly enhanced, giving rise to superconducting $T_c$ of up to hundreds of mK, which is comparable to experimentally reported values \cite{Xia:2024aa,Guo:2024aa}. 
Importantly, due to spin-valley locking in tWSe$_2$, intervalley fluctuations generally acquire an antiferromagnetic character.
The superconducting $T_c$ maxima occur at the van Hove singularities (VHSs) and increases with increasing the displacement field due to the enhanced density of states and IVA fluctuations, ultimately resulting in an IVA ordering. The interplay between superconductivity and IVA order yields a phase diagram consistent with experimental observations \cite{Guo:2024aa}.
At $T_c$, the doubly degenerate superconducting order exhibits $d$-wave-like symmetry, and below $T_c$, self-consistent calculations favor chiral $d \pm id$ states with Chern number $\mathcal{C} = \pm2$.

\textit{Model.---}  Since the $K$ and $K'$ valleys in tWSe$_2$ are related by time-reversal symmetry, we present the $K$-valley (spin up by spin-valley locking) moir\'e Hamiltonian \cite{Wu:2019aa} as follows
\begin{equation}
\mathcal{H}_{K} = 
\begin{pmatrix}
h_{+}(\bm{k}) + U_{+}(\bm{r})  & T (\bm{r}) \\
 T^{\dagger}(\bm{r})   &h_{-}(\bm{k}) + U_{-}(\bm{r})
\end{pmatrix},
\end{equation}
where $h_{\pm}(\bm{k}) = -\hbar^2(\bm{k}-\bm{\kappa}_{\pm})^2/2m^*$ are the effective mass ($m^*=0.43m_0$) approximation of the kinetic energies for the top and bottom WSe$_2$ with $m_0$ being the bare electron mass, and $\bm{\kappa}_{\pm}$ denote the corners of the moiré Brillouin zone (MBZ). The intra- and inter-layer mori\'e potentials are given by 
\begin{equation}
\begin{aligned}
U_{\pm}(\bm{r})   &= 2V\sum_{j =1,3,5}\cos(\bm{G}_j\cdot \bm{r}  \pm \psi) \pm V_z, \\
T (\bm{r}) & = w (1+e^{-i\bm{G}_2\cdot \bm{r}} +e^{-i\bm{G}_3\cdot \bm{r}}),
\end{aligned}
\end{equation}
where $\bm{G}_j$ is the mori\'e reciprocal lattice vector given by rotating $\bm{G}_1 = (4\pi \theta/\sqrt{3}a_0,0)$ counterclockwisely with angle $(j-1)\pi/3$, $a_0 = 3.317$ \AA~is
the lattice constant of monolayer WSe$_2$, and $V_z$ is the layer potential from an applied displacement field. We take model parameters $(V,\psi,w) = (9.0~\text{meV}, 128^{\circ},18~\text{meV})$ from Ref.~\cite{Devakul:2021aa}.

\begin{figure}
 \centering
\includegraphics[width=1\columnwidth]{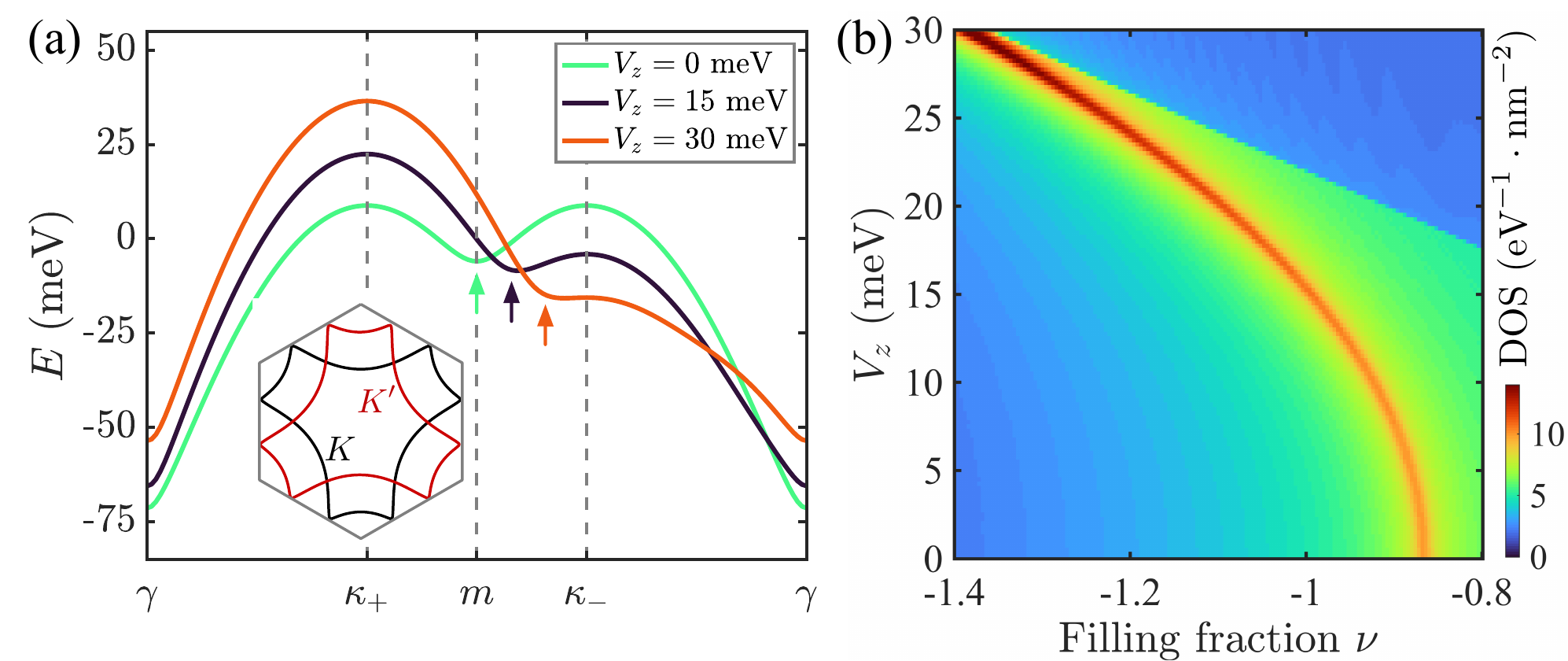}
 \caption{(a) The $K$-valley first mori\'e band along high-symmetry lines of the MBZ. The vertical arrows mark the saddle points for three different values of $V_z$. The inset depicts the Fermi surfaces of $K$ and $K'$ valleys for $V_z = 15$ meV at the VHS. (b) DOS versus the filling fraction $\nu$ and $V_z$. }
 \label{fig:figure1}
\end{figure}

Figure~\ref{fig:figure1}(a) shows the first mori\'e band in $K$ valley for $\theta=5^{\circ}$ at three typical values of $V_z$, revealing a bandwidth of approximately 80 meV. For $V_z=0$ meV, saddle points are found at the $m$ points of the MBZ depicted in the inset. As $V_z$ increases, these saddle points shift away from the $m$ point towards the $\kappa_{-}$ point, as marked by the vertical arrows in Fig.~\ref{fig:figure1}(a).
As shown in Fig.~\ref{fig:figure1}(b), the saddle points give rise to VHSs in the density of states (DOS), which move to higher hole fillings with enhanced magnitude by increasing $V_z$. 

The bare Coulomb interaction in momentum space is
\begin{equation}
H_{c} = \frac{1}{2A}  \sum_{\bm{q}} V_{0}(\bm{q}) \hat{\rho}_{\bm{q}} \hat{\rho}_{-\bm{q}},
\end{equation}
where $A$ denotes the area of the system, $V_{0}(\bm{q}) =(2\pi e^2/\epsilon q ) \tanh(qd_s)$ is the bare Coulomb potential screened by an environmental dielectric constant $\epsilon$ and dual metallic gates placed at a distance $d_s$ away from the sample, $\hat{\rho}_{\bm{q}} = \sum_{\bm{k} \tau} c^{\dagger}_{\bm{k} \tau} c_{\bm{k}+\bm{q} \tau}$ represents the electron density operator, and $c^{\dagger}_{\bm{k}\tau}$ ($c_{\bm{k}\tau}$) is the electron creation (annihilation) operator at valley $\tau$ and wavevector $\bm{k}$. Unless otherwise specified, we choose $\theta=5^{\circ}$, $\epsilon =5$, and $d_s =11$ nm in the following, aligning with experimental conditions in Ref.~\onlinecite{Guo:2024aa}.

\begin{figure}
 \centering
\includegraphics[width=1\columnwidth]{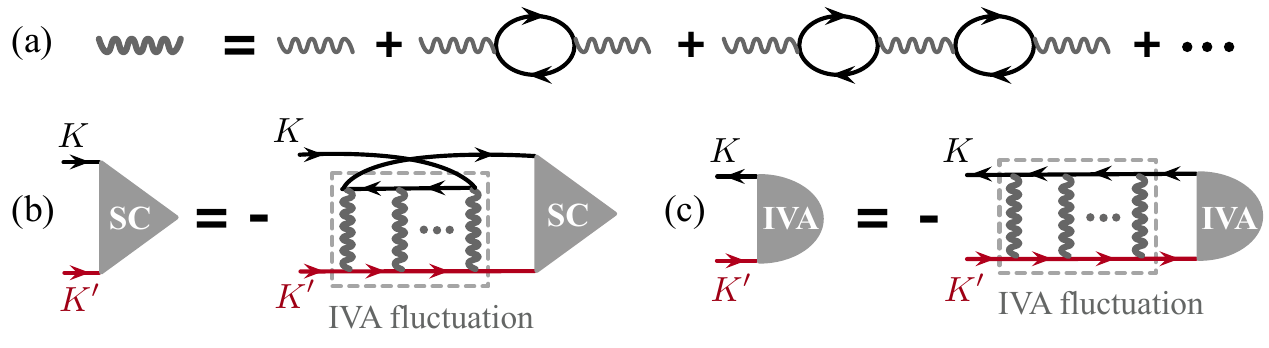}
 \caption{(a) RPA bubble diagram summation, where the light and bold wave lines denote the bare and screened Coulomb interactions, respectively. (b),(c) Diagrammatic representations of the linearized gap equation for (b) pairing instability and	 (c) IVA order. The filled triangle and semicircle denote the SC and IVA order parameters, respectively. The dashed rectangles in (b) and (c) denote series summation of the ladder diagrams, representing IVA fluctuations induced by the RPA screened interaction illustrated in panel (a).}
 \label{fig:figure2}
\end{figure}

\textit{Screened Coulomb interaction.---}  Within the RPA theory \cite{Cea:2021aa,Ghazaryan:2021aa,Cea:2022aa,Sboychakov:2023aa,Long:2024aa,Sboychakov:2024aa,Sboychakov:2025aa}, as diagramatically shown in Fig.~\ref{fig:figure2}(a), the screened Coulomb potential is derived as
\begin{equation}
\label{eq:screenVq}
V^s_{\bm{Q},\bm{Q}'} (\bm{q}) = V_{0} (\bm{q}+\bm{Q}) [1+\sum_{\bm{Q}''} \Pi_{\bm{Q},\bm{Q}''}(\bm{q}) V^s_{\bm{Q}'',\bm{Q}'} (\bm{q})],
\end{equation}
where $\bm{Q}'s$ are mori\'e reciprocal lattice vectors and $\bm{q}$ is the wavevector defined within the MBZ. A compact matrix form of the above equation is $V^s = V_0/(\mathbb{1}-\Pi V_0)$. The static charge polarization function is given by \cite{SM2024}
\begin{equation}
\label{eq:chargepolar}
\begin{aligned}
\Pi_{\bm{Q},\bm{Q}'}(\bm{q}) &= \frac{2}{A}\sum_{mm' \bm{k}}  \frac{f(\epsilon^{K}_{ m\bm{k}})-f(\epsilon^{K}_{ m'\bm{k}+\bm{q}})}{\epsilon^{K}_{m\bm{k}}-\epsilon^{K}_{ m'\bm{k}+\bm{q}}} 
F^{mm'}_{\bm{Q}\bm{Q}'} (\bm{k},\bm{k}+\bm{q}),
\end{aligned}
\end{equation}
where $\epsilon^{K}_{m\bm{k}}$ denotes the $K$-valley electron energy of mori\'e band $m$ at wavevector $\bm{k}$, $f(\epsilon)$ is the Fermi-Dirac distribution function, and the expression of form factor $F^{mm'}_{\bm{Q}\bm{Q}'} (\bm{k},\bm{k}+\bm{q})$ is detailed in the Supplementary Material (SM) \cite{SM2024}. The prefactor 2 in Eq.~(\ref{eq:chargepolar}) accounts for two valleys, each contributing equally to $\Pi(\bm{q})$ \cite{SM2024}. 

We first examine the Kohn-Luttinger mechanism \cite{Kohn:1965aa} of pairing instability driven by screened Coulomb interaction. At low temperatures, the primary contributions to $\Pi$ are from electronic states around the FS. As detailed in the SM \cite{SM2024}, the single-band screened Coulomb interaction can induce pairing instability with critical temperatures of tens of mK. However, when screening effects from remote bands are included, the Coulomb potential is further reduced, failing to induce detectable superconductivity. Consequently, we conclude that the RPA-level Kohn-Luttinger mechanism cannot explain the experimentally observed superconductivity \cite{Guo:2024aa}.

\textit{IVA fluctuations.---}
Motivated by the experimental observation that superconductivity occurs adjacent to a metallic state with a FS reconstruction \cite{Guo:2024aa}, we next explore the effective interaction $V_{\text{eff}}$ mediated by IVA fluctuations . The diagrammatic representation of IVA fluctuation is illustrated in Fig.~\ref{fig:figure2}(b) and (c), where $V_{\text{eff}}$ is computed based on the RPA-screened Coulomb interaction. As detailed in the SM \cite{SM2024}, $V_{\text{eff}}$ is derived as
\begin{equation}
V_{\text{eff}}(\bm{q}_{\text{ex}})= \frac{V_{KK'}(\bm{q}_{\text{ex}})}{\mathbb{1}+ \chi_{KK'}(\bm{q}_{\text{ex}}) V_{KK'}(\bm{q}_{\text{ex}}) },
\label{eq:IVAinteraction}
\end{equation}
where $V_{KK'}(\bm{q}_{\text{ex}})$ denotes the band-projected multi-band screened Coulomb interaction between two electrons from opposite valleys (i.e., opposite spins), $\chi_{KK'}(\bm{q}_{\text{ex}})$ is the intervalley particle-hole susceptibility with nesting wavevector $\bm{q}_{\text{ex}}$. All quantities in Eq.~(\ref{eq:IVAinteraction}) are matrices, whoes elements  are indexed by two independent in- and out-going wavevectors, such as $\bm{k}$ and $\bm{k}'$ \cite{SM2024}.

\begin{figure}
 \centering
\includegraphics[width=\columnwidth]{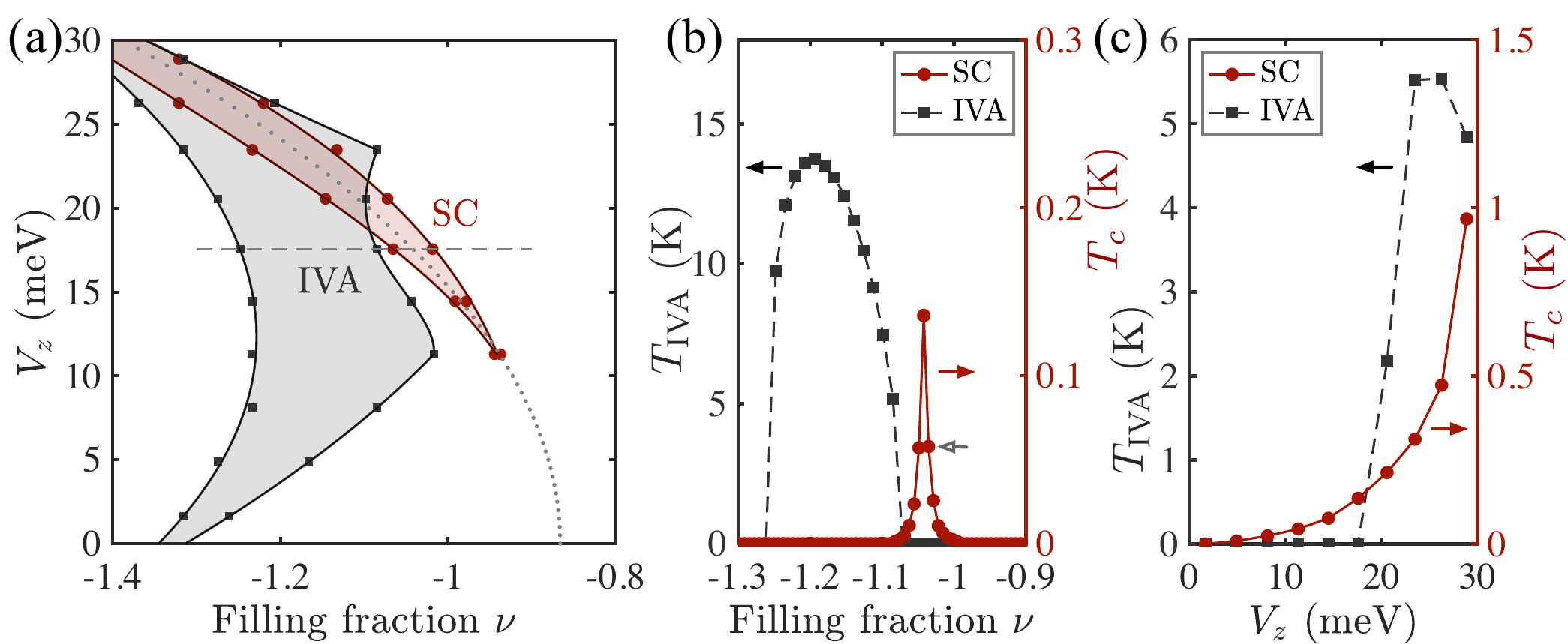}
 \caption{(a) Phase diagram showing the regions of the superconductivity (SC) (red-filled) and IVA order (gray-filled). The dotted line traces the VHSs as functions of  $\nu$ and  $V_z$. (b) SC critical temperature $T_c$ and the IVA critical temperature $T_{\text{IVA}}$ plotted as functions of $\nu$ along the dashed line shown in panel (a), with $V_z = 17.56$ meV. (c) $T_c$ and $T_{\text{IVA}}$ at VHSs, extracted from panel (a) along the doted line. }
 \label{fig:figure3}
\end{figure}

\textit{Linearized gap equation.---}We focus on intervalley electron pairing since its susceptibility diverges at low temperatures due to the time-reversal symmetry, while intravalley susceptibility does not at finite $V_z$ \cite{SM2024}.
The linearized gap equation for superconductivity mediated by IVA fluctuations is schematically depicted in Fig.~\ref{fig:figure2}(b) and given by,
\begin{equation}
\label{eq:LGE}
\Delta_{\bm{k}'} = -\frac{1}{A}\sum_{\bm{k}}V_{\text{eff}}(\bm{k}',-\bm{k}';-\bm{k},\bm{k}) \frac{\tanh(\beta \epsilon^{K}_{\bm{k}}/2)}{2\epsilon^{K}_{\bm{k}}}  \Delta_{\bm{k}},
\end{equation}
where $\Delta_{\bm{k}} $ denotes the order parameter, $\beta = 1/k_BT$ with $k_B$ being the Boltzmann constant, and $\epsilon^{K}_{\bm{k}'}$ is the energy dispersion of the first mori\'e band. 
The effective interaction in the pairing channel is denoted by $V_{\text{eff}}(\bm{k}',-\bm{k}';-\bm{k},\bm{k})$, which is obtained by setting $\bm{q}_{\text{ex}} = \bm{k}+\bm{k}' $ in Eq.~(\ref{eq:IVAinteraction}) \cite{SM2024}.
The $T_c$ is determined by equaling the largest eigenvalue of Eq.~({\ref{eq:LGE}}) to 1, with the pairing symmetry encoded in the eigenvector(s). 
As detailed in the SM \cite{SM2024}, the numerical solution of Eq.~(\ref{eq:LGE}) is performed using a two-step discretization scheme, with an energy truncation set by the bandwidth of the first moiré band. This approach yields smooth variations of $T_c$ with an accuracy of a few mK.

In analogy to the theory of superconductivity mediated by antiferromagnetic fluctuations in cuprates \cite{Scalapino:1986aa,Scalapino:1987aa,Millis:1990aa,Moriya:1990aa},  the IVA fluctuations in this system may give rise to IVA ordering, which competes with superconductivity. The linearized gap equation for IVA is depicted in Fig.~\ref{fig:figure2}(c), which represents a generalized Stoner instability at finite wavevector $\bm{q}_{\text{ex}}$. The critical temperature $T_{\text{IVA}}$ is determined by solving 
\begin{equation}
\eta_{\bm{k}'} = - \frac{1}{A} \sum_{\bm{k}}V_{\text{eff}}(\bm{k}',\bm{k}-\bm{q}_{\text{ex}};\bm{k}'-\bm{q}_{\text{ex}},\bm{k}) \chi_{KK'}(\bm{k},\bm{q}_{\text{ex}})\eta_{\bm{k}},
\label{eq:IVAlge}
\end{equation}
where $\eta_{\bm{k}}$ denotes the IVA order parameter for the $\bm{q}_{ex}$ channel, and $ \chi_{KK'}(\bm{k},\bm{q}_{ex})$ is the corresponding $\bm{k}$-resolved intervalley particle-hole susceptibility. Similar to the case for pairing instability, $T_{\text{IVA}}$ is determined by equaling the largest eigenvalue of Eq.~({\ref{eq:IVAlge}}) to 1. In this study, we focus on the four typical commensurate intervalley orders, including the intervalley coherent state with $\bm{q}_{ex}=\bm{\gamma}$, $\bm{m}$, $\bm{\kappa}_{\pm}$, respectively. Among them, we find that the IVA order associated with $\bm{q}_{ex}=\bm{\kappa}_{+}$ exhibits the strongest instability when $V_z$ exceeds $12$ meV \cite{SM2024}. Hereafter, unless otherwise specified, the IVA order refers to the strongest case with $\bm{q}_{ex}=\bm{\kappa}_{+}$, a $120^{\circ}$ antiferromagnetic order with  $\sqrt{3}\times\sqrt{3}$ magnetic unit cell\cite{Pan:2020aa,Qiu:2023aa,Li:2024aa}.

\textit{Phase diagram.---} Figure~\ref{fig:figure3}(a) shows the phase diagram of SC and IVA state as functions of $\nu$ and $V_z$, where the SC emerges in the vicinity of VHSs, as indicated by the red shaded regime. The IVA sate, represented by the gray region, dominates over the SC state in the overlapping regime when $V_z$ exceeds approximately $20$ meV. Consequently, the SC state persists only within a narrow phase regime around the VHSs at $\nu = -1$, consistent with experiments showing SC emerging at finite displacement fields and being suppressed by antiferromagnetic order as the field increases \cite{Guo:2024aa}. Figure~\ref{fig:figure3}(b) shows the $T_c$ and $T_{\text{IVA}}$ as functions of $\nu$ for a fixed value of $V_z=17.56$ meV, where the $T_{\text{IVA}}$ exhibits a dome-shaped feature, reaching a maximum of $\sim$14 K at $\nu = -1.2$. By decreasing $|\nu|$, the IVA state is gradually suppressed, and the SC emerges once the IVA sate is completely suppressed at $\nu = -1.1$. The $T_c$ exhibits a peak at $\nu=-1.05$, corresponding to the VHS. 
Figure~\ref{fig:figure3}(c) extracts the $T_c$ and $T_{\text{IVA}}$ along the VHS trajectory shown in Fig.~\ref{fig:figure3}(a).
The $T_c$ closely follows the DOS at the van Hove filling, increasing monotonously with $V_z$ due to the enhancement of the DOS (see Fig.~\ref{fig:figure1}(b)), which indicates that the superconductivity is in the weak-coupling regime. When $V_z$ exceeds $\sim$20 meV, the IVA order emerges and dominates over the pairing instability.  The $T_c$ near the IVA ordering onset is around 0.15 K, on the same order of magnitude as the experimental value of $0.4$ K \cite{Guo:2024aa}.

\begin{figure}
 \centering
\includegraphics[width=1\columnwidth]{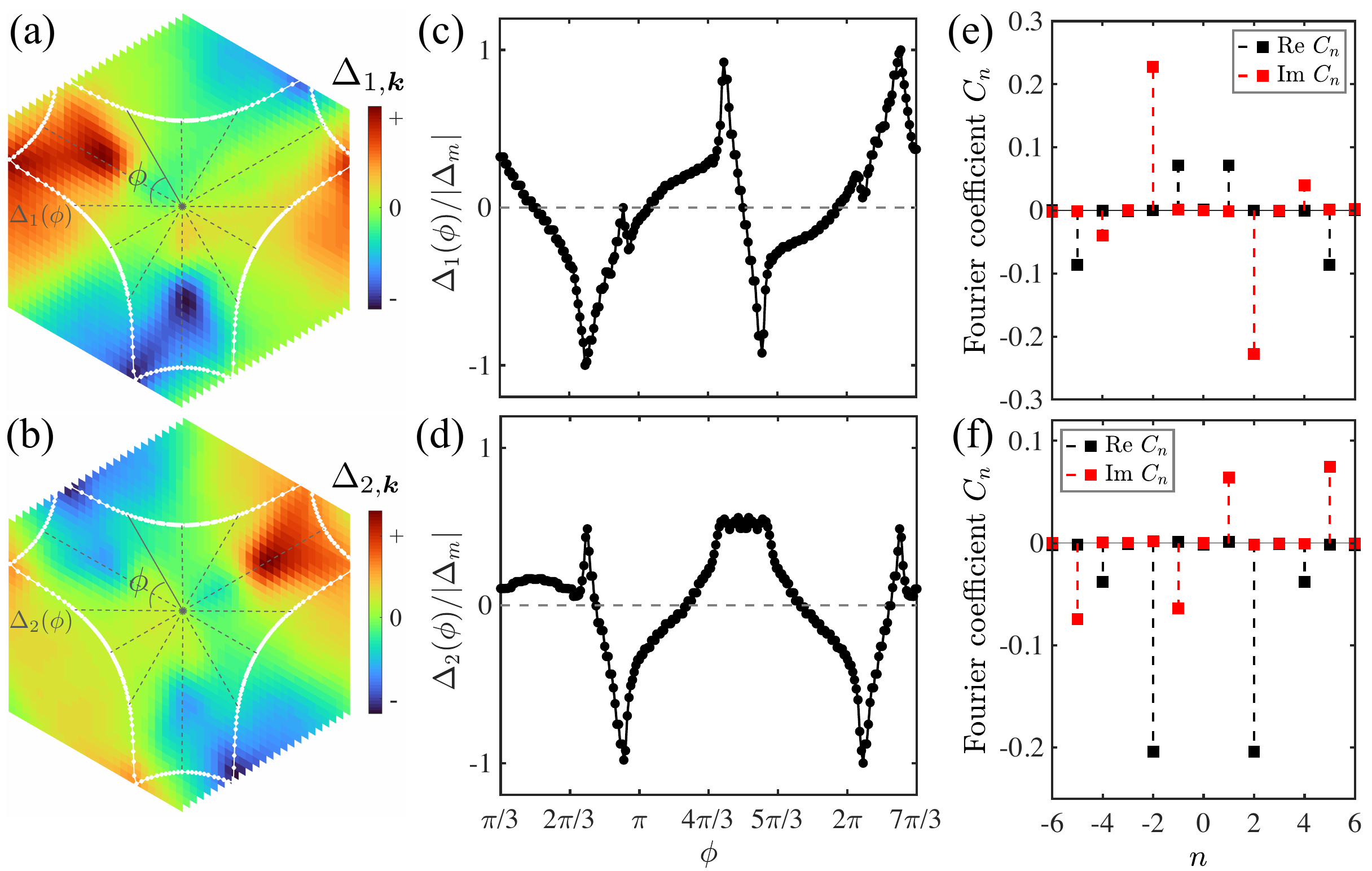}
 \caption{(a),(b) $\bm{k}$-space structure of the doubly-degenerate order parameters $\Delta_{1,\bm{k}}$ and $\Delta_{2,\bm{k}}$ at the phase point indicated by the open arrow in Fig.~\ref{fig:figure3}(b). 
The while lines denote the FS, and the overlaid dots mark the sampling points along the FS, which are equally spaced in angular coordinate $\phi$ with respect to the MBZ center.
(c),(d) Angular dependence of the FS-sampled profiles $\Delta_{1,2}(\phi)$, extracted from (a) and (b), and normalized by their respective maximum magnitudes $|\Delta_{m}|$.
 (e),(f) Fourier decompositions of $\Delta_{1,2}(\phi)$, where $C_n$ denotes the Fourier coefficient of the $n$-th angular momentum channel.
}
 \label{fig:figure4}
\end{figure}

\textit{Pairing symmetry.--} We now examine the pairing order parameter $\Delta_{\bm{k}}$ near the VHS for $V_z= 17.56$ meV, and find that pairing is doubly degenerate at $T_c$, characterized by two orthogonal order parameters, $\Delta_{1,\bm{k}}$ and $\Delta_{2,\bm{k}}$, as plotted in Figs.~\ref{fig:figure4}(a) and \ref{fig:figure4}(b). 
The extracted angular-dependence of $\Delta_{1,2}(\phi)$ along the FS are shown in Figs.~\ref{fig:figure4}(c) and (d), each exhibiting four nodes within a $2\pi$ period, suggesting a $d$-wave-like pairing symmetry. To quantify the analysis, we perform Fourier decomposition 
$
\Delta(\phi) = \sum_{n} C_n e^{in \phi},
$
where $C_n$ denotes the Fourier coefficient of angular momentum channel $n$. 
As shown in Figs.~\ref{fig:figure4}(e) and (f), the dominant contributions to $\Delta_{1,2}(\phi)$ are from the $d$-wave channel with $n = \pm 2$. In particular, $\Delta_{1} (\phi)\propto  \sin2\phi $ and $\Delta_{2} (\phi)\propto  \cos2\phi $, suggesting $d_{x^2-y^2}$ and $d_{xy}$ symmetries, respectively. The doubly degenerate states transform as the $E$ irreducible representation of $\hat{C}_{3}$ point group for tWSe$_2$ at finite $V_z$. The eigenvectors of $E$ representation comprise mixtures of $(p_x, p_y)$, $(d_{x^2-y^2}, d_{xy})$, and higher harmonics, leading to the coexistence of $d$-wave, $p$-wave ($n=\pm 1$), and $g$-wave ($n=\pm 4$) channels, as illustrated in Figs.~\ref{fig:figure4}(e) and (f). The coexistence of even-parity (spin-singlet) and odd-parity (spin-triplet) in the superconducting order parameter results from the absence of inversion symmetry.

\begin{figure}
 \centering
\includegraphics[width=\columnwidth]{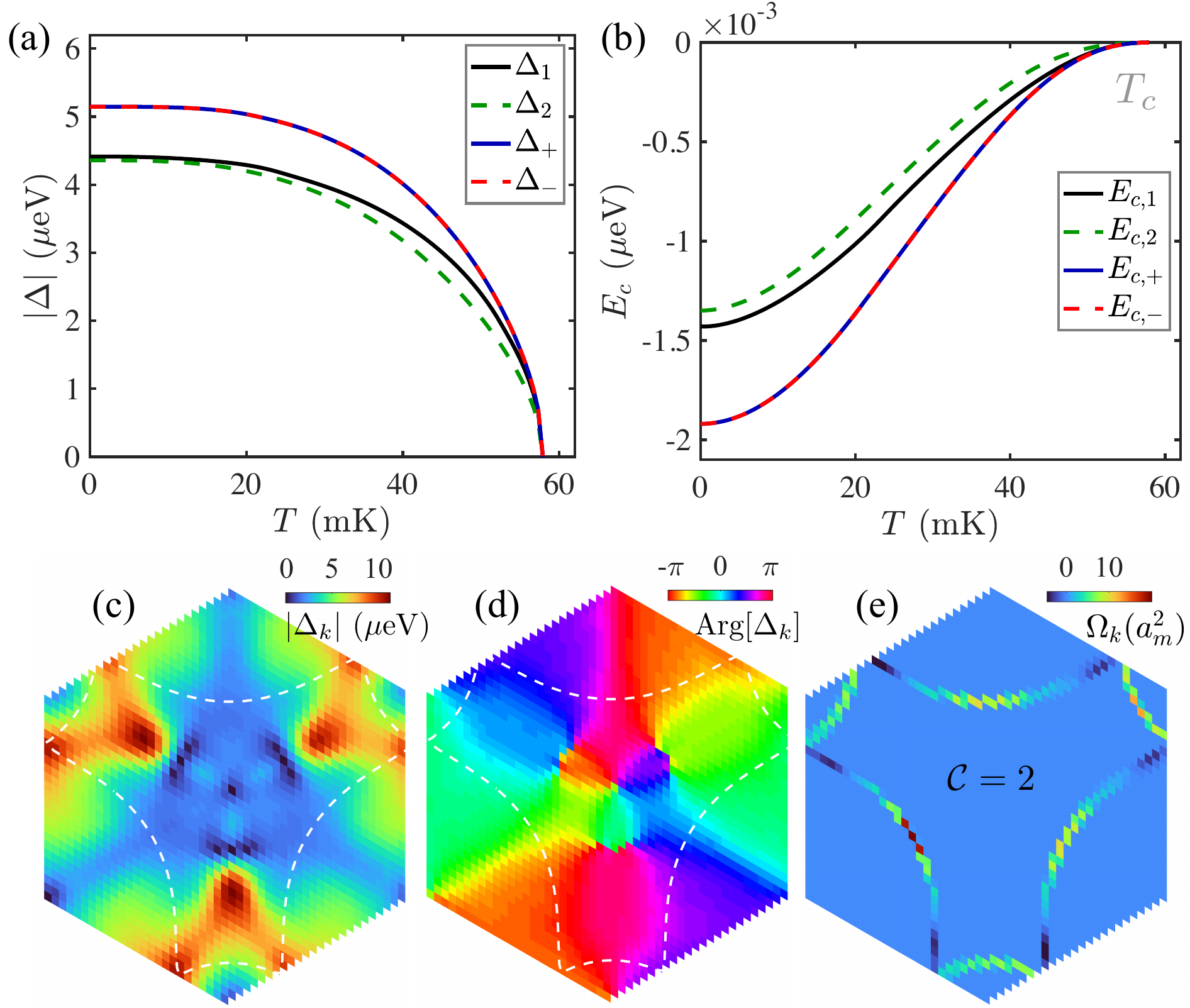}
 \caption{(a) Temperature dependences of the MBZ-averaged magnitudes of the nematic ($\Delta_{1,2}(\bm{k})$) and chiral ($\Delta_{\pm}(\bm{k}) = \Delta_{1}(\bm{k}) \pm i \Delta_{2}(\bm{k})$) superconducting order parameters. (b) Temperature dependences of the superconducting condensation energies ($E_c$) of the nematic and chiral states. (c),(d) Magnitude and phase of $\Delta_{+}(\bm{k})$ at zero temperature, with the dashed curve denoting the FS.
 (d) Berry curvature $\Omega_{\bm{k}}$ for $\Delta_{+}(\bm{k})$ pairing state, where $a_m=a_0/\theta$ is the mori\'e period. These results are obtained by solving the self-consistent gap equation
 with $V_z = 17.56$ meV and $\nu = \nu_{\text{VHS}} + 0.0067$, as marked by the open arrow in Fig.~\ref{fig:figure3}(b). }
 \label{fig:figure5}
\end{figure}

\textit{Chiral superconductivity.--}
We further investigate the 
superconducting ground state at temperatures below $T_c$, focusing on the competition between the nematic superconducting states $\Delta_{1,2}$ and the chiral superconducting states $\Delta_{\pm}(\bm{k}) = \Delta_{1}(\bm{k})\pm i \Delta_{2}(\bm{k})$. 
We solve the self-consistent gap equations for the nematic and chiral pairing states at different temperatures, using $V_z = 17.56$ meV near the VHS as a representative example. 
The temperature dependences of the averaged magnitudes of the order parameters over the MBZ are depicted in  Fig.~\ref{fig:figure5}(a). At $T_c$, the nematic and chiral order parameters are degenerate, consistent with the results from solving the linearized gap equation. Below $T_c$, the degeneracy between the two nematic sates is lifted due to contributions from higher-order (sixth order and beyond) terms in the self-consistent gap equation \cite{Fu:2014aa}. Moreover, $|\Delta_{\pm}|$ attain larger values than $|\Delta_{1,2}|$, indicating that the chiral states are energetically more favorable. 
This is confirmed by the condensation energy $E_c$ calculations, as shown in Fig.~\ref{fig:figure5}(b), where $E_{c,\pm}$ for the chiral states are lower than $E_{c,1,2}$ for the nematic states. Figures~\ref{fig:figure5}(c) and (d) display the magnitude and phase of $\Delta_{+}(\bm{k})$ at zero temperature, where the presence of four winding phase points (each covering 0 or $2\pi$) along the FS suggests a chiral $d$-wave pairing state. The chiral superconducting states spontaneously breaks time-reversal symmetry and typically has nontrivial topological properties. Figure~\ref{fig:figure5}(e) shows the Berry curvature $\Omega_{\bm{k}}$ calculated using the Bogoliubov-de Gennes Hamiltonian, where $\Omega_{\bm{k}}$ emerges around the FS due to the small superconducting gap. The resulting Chern number $\mathcal{C} = 2$ is consistent with the topological chiral $d+id$ superconducting sates \cite{Nandkishore:2012aa,Qin:2023aa}.

\textit{Discussion.---} 
This study proposes a purely electronic mechanism for the superconductivity observed recently in tWSe$_2$ at $\theta=5^{\circ}$ \cite{Xia:2024aa,Guo:2024aa}.
While multiband RPA-screened Coulomb interaction alone fails to induce pairing, IVA fluctuations significantly enhance the pairing instability, leading to superconductivity near the VHS at $\nu = -1$ with $T_c \sim 0.12$ K—consistent with experiments at $\theta = 5^{\circ}$ \cite{Guo:2024aa}. The superconducting state exhibits doubly degenerate $d$-wave-like pairing, favoring chiral $d \pm id$ states below $T_c$ with Chern number $\mathcal{C} = \pm 2$.
Below $T_c$, a uniform system will spontaneously evolve into one chiral state, thereby breaking the time-reversal symmetry. Inhomogeneities may lead to opposite chiral domains separated by interfaces hosting chiral edge states.

For the IVA order, the present study only considers four representative commensurate states with wavevectors at the high-symmetry points of the MBZ. The incommensurate IVA ordering states may also compete with superconductivity, which calls further investigation, such as through the functional renormalization group calculations \cite{Qin:2023aa,Fischer:2025aa}.
We emphasize that the pairing instability calculation includes contributions from IVA fluctuations at all wavevectors.
The present theory does not incorporate interaction-induced band renormalization, which could modify the phase diagram shown in Fig.~\ref{fig:figure3}(c). This renormalization may become dramatic for systems with smaller $\theta$, where the electronic correlations are significantly enhanced, as evidenced by the emergence of correlated insulator states \cite{Wang:2020aa,Xia:2024aa}, rendering reliance on perturbative calculations invalid. 
Our theory is particularly suited for tWSe$_2$ at larger twist angles, such as $\theta = 5^{\circ}$, where the Fermi surface instability analysis yields a theoretical phase diagram featuring competing orders in qualitative agreement with experiment.

\textit{Acknowledgement.---}
WQ and WXQ contributed equally to this work. This work is supported by National Natural Science Foundation of China (Grant Nos.12474134 and 1227433)

\textit{Note added.---}During the preparation of this manuscript, we became aware of a related preprint posted on arxiv \cite{Guerci:2024aa}.

\end{document}


\title{Supplemental Material for ``Topological Chiral Superconductivity Mediated by Intervalley Antiferromagnetic Fluctuations in Twisted Bilayer WSe$_2$"}
\author{Wei Qin}
\email{qinwei5@ustc.edu.cn}
\affiliation{Department of Physics, University of Science and Technology of China, Hefei, Anhui 230026,China}
\author{Wen-Xuan Qiu}
\thanks{These authors contributed equally to this work.}
\affiliation{School of Physics and Technology, Wuhan University, Wuhan 430072, China}
\author{Fengcheng Wu}
\email{wufcheng@whu.edu.cn}
\affiliation{School of Physics and Technology, Wuhan University, Wuhan 430072, China}
\date{\today}

\maketitle

\section{Charge polarization function}
The random phase approximation (RPA) represents a summation of charge bubble diagrams connected via bare Coulomb interaction (see Fig.~2(a) in the main text). In the static limit, the charge bubble is characterized by the static charge polarization function. Because (1) the wave vector difference between two valleys significantly exceeds the size of mori\'e Brillouin zone (MBZ) and (2) the two valleys have opposite spin polarizations, the valley-flipping effects caused by bare Coulomb interaction can be neglected. Therefore, the static charge polarization function is diagonal in the valley subspace and can be expressed in the plane-wave basis as
\begin{equation}
\Pi^{\tau \tau}_{\bm{Q},\bm{Q}'}(\bm{q}) = \frac{1}{\beta A} \sum_{i\omega_n}\sum_{\tau \bm{k}}  \sum_{\bm{G} \bm{G}'}\sum_{ll'} \mathcal{G}^{\tau}_{l, \bm{G};l', \bm{G}'}(i\omega_n,\bm{k}) \mathcal{G}^{\tau}_{l',\bm{G}'+\bm{Q}';l, \bm{G}+\bm{Q}}(i\omega_n,\bm{k}+\bm{q}),
\label{eq:ChargPF}
\end{equation}
where $\tau$ is the valley index, $\beta=1/k_BT$, $A$ denotes the area of the system, $\omega_n = (2n+1)\pi$ is the fermionic Matsubara frequency, $l$ is the layer index, $\bm{G}$ and $\bm{Q}$ are mori\'e reciprocal lattice vectors. By expanding the single-particle Green's function in the band representation, we have 
\begin{equation}
\mathcal{G}^{\tau} (i\omega_n,\bm{k}) = \sum_{m} \frac{|\tau m\bm{k}\rangle \langle \tau m\bm{k} |}{i\omega_n-\epsilon^{\tau}_{ m \bm{k}}} =  \sum_{m} \frac{\rho^{\tau}_{ m \bm{k}}}{i\omega_n-\epsilon^{\tau}_{ m \bm{k}}} ,
\label{eq:Greenfun}
\end{equation}
where $\epsilon^{\tau}_{ m  \bm{k}}$ is the electron energy specified by valley $\tau$, band $m$, and wavevector $\bm{k}$, and $\rho^{\tau}_{ m \bm{k}} = |\tau m\bm{k}\rangle \langle \tau m\bm{k} |$ denotes the density matrix defined by the state $|\tau m\bm{k}\rangle$. Here $|\tau m\bm{k}\rangle$ is the periodic part of the Bloch wave function.  By substituting Eq.~(\ref{eq:Greenfun}) into Eq.~(\ref{eq:ChargPF}),
\begin{equation}
\begin{aligned}
\Pi_{\bm{Q},\bm{Q}'}^{\tau \tau}(\bm{q}) &= \frac{1}{\beta A} \sum_{i\omega_n}\sum_{ \bm{k}}  \sum_{\bm{G} \bm{G}'}\sum_{ll'} \sum_{ mm'} \frac{[\rho^{\tau}_{m\bm{k}}]_{l,\bm{G};l',\bm{G}'}}{i\omega_n-\epsilon^{\tau}_{m \bm{k}}}  \frac{ [\rho^{\tau}_{m'\bm{k} +\bm{q}}]_{l', \bm{G}'+\bm{Q}';l,\bm{G}+\bm{Q}} }{i\omega_n-\epsilon^{\tau}_{m' \bm{k}+\bm{q}}} \\
& = \frac{1}{ A} \sum_{ mm' \bm{k}}  \frac{f(\epsilon^{\tau}_{m \bm{k}})-f(\epsilon^{\tau}_{m' \bm{k}+\bm{q}})}{\epsilon^{\tau}_{m \bm{k}}-\epsilon^{\tau}_{m' \bm{k}+\bm{q}}} 
\sum_{ll' \bm{G} \bm{G}'} [\rho^{\tau}_{m\bm{k}}]_{l,\bm{G};l',\bm{G}'}  [\rho^{\tau}_{m'\bm{k} +\bm{q}}]_{l', \bm{G}'+\bm{Q}';l,\bm{G}+\bm{Q}} \\
& = \frac{1}{ A} \sum_{ mm' \bm{k}}  \frac{f(\epsilon^{\tau}_{m \bm{k}})-f(\epsilon^{\tau}_{m' \bm{k}+\bm{q}})}{\epsilon^{\tau}_{m \bm{k}}-\epsilon^{\tau}_{m' \bm{k}+\bm{q}}} 
F^{\tau m,\tau m'}_{\bm{Q},\bm{Q}'} (\bm{k},\bm{k}+\bm{q}),
\end{aligned}
\label{eq:ChargPF1}
\end{equation}
where the corresponding form factor is defined as
\begin{equation}
F^{\tau m,\tau m'}_{\bm{Q},\bm{Q}'} (\bm{k},\bm{k}+\bm{q}) = \sum_{ll' \bm{G} \bm{G}'} [\rho^{\tau}_{m\bm{k}}]_{l,\bm{G};l',\bm{G}'}  [\rho^{\tau}_{m'\bm{k} +\bm{q}}]_{l', \bm{G}'+\bm{Q}';l,\bm{G}+\bm{Q}} .
\end{equation}
Here we show that $\Pi^{\tau \tau}(\bm{q})$ possesses several properties. First, from Eq.~(\ref{eq:ChargPF1}),  we have
\begin{equation}
\begin{aligned}
\left[ \Pi_{\bm{Q}',\bm{Q}}^{\tau \tau}(\bm{q}) \right]^*&= \frac{1}{ A} \sum_{ mm' \bm{k}}  \frac{f(\epsilon^{\tau}_{m \bm{k}})-f(\epsilon^{\tau}_{m' \bm{k}+\bm{q}})}{\epsilon^{\tau}_{m \bm{k}}-\epsilon^{\tau}_{m' \bm{k}+\bm{q}}} 
\sum_{ll' \bm{G} \bm{G}'} [\rho^{\tau}_{m\bm{k}}]^*_{l,\bm{G};l',\bm{G}'}  [\rho^{\tau}_{m'\bm{k} +\bm{q}}]^*_{l', \bm{G}'+\bm{Q};l,\bm{G}+\bm{Q}'} \\
&= \frac{1}{ A} \sum_{ mm' \bm{k}}  \frac{f(\epsilon^{\tau}_{m \bm{k}})-f(\epsilon^{\tau}_{m' \bm{k}+\bm{q}})}{\epsilon^{\tau}_{m \bm{k}}-\epsilon^{\tau}_{m' \bm{k}+\bm{q}}} 
\sum_{ll' \bm{G} \bm{G}'} [\rho^{\tau}_{m\bm{k}}]_{l',\bm{G}'; l,\bm{G}}  [\rho^{\tau}_{m'\bm{k} +\bm{q}}]_{l,\bm{G}+\bm{Q}'; l', \bm{G}'+\bm{Q}} \\
&=\Pi^{\tau \tau}_{\bm{Q},\bm{Q}'}(\bm{q}) 
\end{aligned}.
\end{equation}
In the above equation, we have used the relation $\rho^{\tau}_{m\bm{k}}=[\rho^{\tau}_{m\bm{k}}]^{\dagger}$. The above equation implies $\Pi^{\tau \tau}(\bm{q}) = [\Pi^{\tau \tau}(\bm{q})]^{\dagger} $ is a Hermitian matrix.
Secondly, we have
\begin{equation}
\begin{aligned}
\Pi^{\tau \tau}_{-\bm{Q}',-\bm{Q}}(-\bm{q}) &= \frac{1}{ A} \sum_{ mm' \bm{k}}  \frac{f(\epsilon^{\tau}_{m \bm{k}})-f(\epsilon^{\tau}_{m' \bm{k}-\bm{q}})}{\epsilon^{\tau}_{m \bm{k}}-\epsilon^{\tau}_{m' \bm{k}-\bm{q}}} 
\sum_{ll' \bm{G} \bm{G}'} [\rho^{\tau}_{m\bm{k}}]_{l,\bm{G};l',\bm{G}'}  [\rho^{\tau}_{m'\bm{k} -\bm{q}}]_{l', \bm{G}'-\bm{Q};l,\bm{G}-\bm{Q}'} \\
&= \frac{1}{ A} \sum_{ mm' \bm{k}}  \frac{f(\epsilon^{\tau}_{m \bm{k}+\bm{q}})-f(\epsilon^{\tau}_{m' \bm{k}})}{\epsilon^{\tau}_{m \bm{k}+\bm{q}}-\epsilon^{\tau}_{m' \bm{k}}} 
\sum_{ll' \bm{G} \bm{G}'} [\rho^{\tau}_{m\bm{k}+\bm{q}}]_{l,\bm{G};l',\bm{G}'}  [\rho^{\tau}_{m'\bm{k} }]_{l', \bm{G}'-\bm{Q};l,\bm{G}-\bm{Q}'} \\
&= \frac{1}{ A} \sum_{ mm' \bm{k}}  \frac{f(\epsilon^{\tau}_{m' \bm{k}+\bm{q}})-f(\epsilon^{\tau}_{m \bm{k}})}{\epsilon^{\tau}_{m' \bm{k}+\bm{q}}-\epsilon^{\tau}_{m \bm{k}}} 
\sum_{ll' \bm{G} \bm{G}'} [\rho^{\tau}_{m'\bm{k}+\bm{q}}]_{l,\bm{G}'+\bm{Q}';l',\bm{G}+\bm{Q}}  [\rho^{\tau}_{m\bm{k} }]_{l', \bm{G};l,\bm{G}'} \\
& =\Pi^{\tau \tau}_{\bm{Q},\bm{Q}'}(\bm{q}),
\end{aligned},
\end{equation}
which indicates that charge polarization function is inversion symmetric. Thirdly, it can be demonstrate that $\Pi^{\tau \tau}_{\bm{Q},\bm{Q}'}(\bm{q}) = \Pi^{\tau \tau}_{\hat{C}_3 \bm{Q},\hat{C}_3\bm{Q}'}(\hat{C}_3\bm{q})$ by recognizing that the system preserves three-fold rotational symmetry. 
Finally, time-reversal symmetry (TRS) imposes the relations
\begin{equation}
\begin{aligned}
\epsilon^{K}_{m, \bm{k}} &=\epsilon^{K'}_{m, -\bm{k}}, \\
[\rho^{K}_{m,\bm{k}}]_{l,\bm{G};l',\bm{G}'}  & = [\rho^{K'}_{m,-\bm{k}}]^*_{l,-\bm{G};l',-\bm{G}'}.
\end{aligned}
\label{eq:TRS}
\end{equation}
Therefore, we have
\begin{equation}
\begin{aligned}
\Pi^{K K}_{\bm{Q},\bm{Q}'}(\bm{q}) & = \frac{1}{ A} \sum_{ mm' \bm{k}}  \frac{f(\epsilon^{K}_{m, \bm{k}})-f(\epsilon^{K}_{m', \bm{k}+\bm{q}})}{\epsilon^{K}_{m, \bm{k}}-\epsilon^{K}_{m', \bm{k}+\bm{q}}} 
\sum_{ll' \bm{G} \bm{G}'} [\rho^{K}_{m,\bm{k}}]_{l,\bm{G};l',\bm{G}'}  [\rho^{K}_{m',\bm{k} +\bm{q}}]_{l', \bm{G}'+\bm{Q}';l,\bm{G}+\bm{Q}}\\
& = \frac{1}{ A} \sum_{ mm' \bm{k}}  \frac{f(\epsilon^{K'}_{m, -\bm{k}})-f(\epsilon^{K'}_{m', -\bm{k}-\bm{q}})}{\epsilon^{K'}_{m, -\bm{k}}-\epsilon^{K'}_{m', -\bm{k}-\bm{q}}} 
\sum_{ll' \bm{G} \bm{G}'} [\rho^{K'}_{m,-\bm{k}}]^*_{l,-\bm{G};l',-\bm{G}'}  [\rho^{K'}_{m',-\bm{k} -\bm{q}}]^*_{l', -\bm{G}'-\bm{Q}';l,-\bm{G}-\bm{Q}}\\
& = \frac{1}{ A} \sum_{ mm' \bm{k}}  \frac{f(\epsilon^{K'}_{m, \bm{k}})-f(\epsilon^{K'}_{m', \bm{k}-\bm{q}})}{\epsilon^{K'}_{m, \bm{k}}-\epsilon^{K'}_{m', \bm{k}-\bm{q}}} 
\sum_{ll' \bm{G} \bm{G}'} [\rho^{K'}_{m,\bm{k}}]^*_{l,\bm{G};l',\bm{G}'}  [\rho^{K'}_{m',\bm{k} -\bm{q}}]^*_{l', \bm{G}'-\bm{Q}';l,\bm{G}-\bm{Q}}\\
&=[\Pi^{K' K'}_{-\bm{Q},-\bm{Q}'}(-\bm{q}) ]^*.
\end{aligned}
\end{equation}
These properties of the charge polarization function are summarized as follows:
\begin{equation}
\begin{aligned}
 \Pi^{\tau \tau}_{\bm{Q},\bm{Q}'}(\bm{q}) & =  [\Pi_{\bm{Q}',\bm{Q}}^{\tau \tau} (\bm{q})]^*,  \\ 
 \Pi^{\tau \tau}_{\bm{Q},\bm{Q}'}(\bm{q}) & = \Pi^{\tau \tau}_{-\bm{Q}',-\bm{Q}}(-\bm{q}), \\
  \Pi^{\tau \tau}_{\bm{Q},\bm{Q}'}(\bm{q}) & = \Pi^{\tau \tau}_{\hat{C}_3 \bm{Q},\hat{C}_3\bm{Q}'}(\hat{C}_3\bm{q}), \\
  \Pi^{KK}_{\bm{Q},\bm{Q}'}(\bm{q}) &= [\Pi^{K'K'}_{-\bm{Q},-\bm{Q}'}(-\bm{q}) ]^*.
\end{aligned}
\label{eq:relations}
\end{equation}
Based on these relations, we have $  \Pi^{KK}_{\bm{Q},\bm{Q}'}(\bm{q}) =  [\Pi^{K'K'}_{\bm{Q}',\bm{Q}}(\bm{q}) ]^* = \Pi^{K'K'}_{\bm{Q},\bm{Q}'}(\bm{q})$, indicating that the two valleys give equal contributions to the total polarization function. Therefore, we have 
\begin{equation}
\begin{aligned}
\Pi_{\bm{Q},\bm{Q}'}(\bm{q}) &=  \Pi^{KK}_{\bm{Q},\bm{Q}'}(\bm{q}) +  \Pi^{K'K'}_{\bm{Q},\bm{Q}'}(\bm{q})  =
\frac{2}{A}\sum_{mm' \bm{k}}  \frac{f(\epsilon^{K}_{ m\bm{k}})-f(\epsilon^{K}_{ m'\bm{k}+\bm{q}})}{\epsilon^{K}_{m\bm{k}}-\epsilon^{K}_{ m'\bm{k}+\bm{q}}} 
F^{Km,Km'}_{\bm{Q},\bm{Q}'} (\bm{k},\bm{k}+\bm{q}),
\end{aligned}
\label{eq: chargeP}
\end{equation}
where the prefactor 2 accounts for two valleys. The above equation recovers Eq.~(5) in the main text.

\section{Screened Coulomb interaction} 
\label{sec:RPA_screening}
The RPA-level screened Coulomb potential $V^s(\bm{q})$, derived from Fig.~2(a) in the main text, can be expressed in a compact form as
\begin{equation}
V^s(\bm{q}) = \frac{V_0(\bm{q})}{\mathbb{1}-\Pi(\bm{q}) V_0(\bm{q})}.
\label{eq: screenedCP}
\end{equation}
Here the bare Coulomb potential $V_0(\bm{q})$, $V^s(\bm{q})$, and $\Pi(\bm{q})$ are matrices defined in the basis of mori\'e reciprocal lattice vectors. In particular, $V_0(\bm{q})$ is a diagonal matrix with elements given by $[V_0(\bm{q})]_{\bm{Q},\bm{Q}'} = V_0(\bm{q}+\bm{Q})\delta_{\bm{Q},\bm{Q}'} $, which naturally satisfies the relations similar to those outlined in Eq.~(\ref{eq:relations}). Therefore, it can be readily demonstrated that the $V^{s}_{\bm{Q},\bm{Q}'}(\bm{q})$ also complies with the relations similar to those summarized in Eq.~(\ref{eq:relations}).

We first examine the possibility of superconductivity induced by the screened Coulomb interaction via Kohn-Luttinger mechanism \cite{Kohn:1965aa}. This study primarily considers inter-valley electron pairing, as the intra-valley pairing generally involves additional energy costs arising from trigonal wrapping effect of the single-particle band structure. The inter-valley pairing channel interaction is therefore given by 
\begin{equation}
H_{in} = \sum_{\bm{k}\bm{k}'} \sum_{\bm{Q},\bm{Q}'} V^s_{\bm{Q}\bm{Q}'}(\bm{k}-\bm{k}') \sum_{l l'\bm{G}\bm{G}'}  \psi^{\dagger}_{K,l}(\bm{k}'+\bm{G}-\bm{Q}) \psi^{\dagger}_{K',l'}(-\bm{k}'+\bm{G}'+\bm{Q}')  \psi_{K',l'}(-\bm{k}+\bm{G}') \psi_{K,l}(\bm{k}+\bm{G}),
\label{eq:planewaveHp}
\end{equation}
where $\psi_{K,l}(\bm{k}+\bm{G})$ denotes the annihilation operator for the electron from valley $K$ and layer $l$, with plane wavevector $\bm{k}+\bm{G}$. The relation between operators in the plane-wave and band representations is provided as
\begin{equation}
\psi^{\dagger}_{\tau l}(\bm{k}+\bm{G}) = \sum_{m} \langle \tau, m, \bm{k} |\tau, l, \bm{G} \rangle c_{\tau m}^{\dagger}(\bm{k}),
\label{eq:operator_rel}
\end{equation}
where $c_{\tau m}^{\dagger}(\bm{k})$ denotes the creation operator for the electron from valley $\tau$ and mori\'e band $m$, with wavevector $\bm{k} \in \text{MBZ}$. By substituting Eq.~(\ref{eq:operator_rel}) into Eq.~(\ref{eq:planewaveHp}) and keeping $m=1$, the first mori\'e band-projected pairing Hamiltonian is
\begin{equation}
H_p = \sum_{\bm{k}\bm{k'}}V_b( \bm{k}',\bm{k}) c_{K}^{\dagger} (\bm{k}') c_{K'}^{\dagger}(-\bm{k}')  c_{K'}(-\bm{k})  c_{K}(\bm{k}). 
\end{equation}
The pairing interaction in the above equation is derived as
\begin{equation}
\begin{aligned}
V_b( \bm{k}',\bm{k}) &= \sum_{\bm{Q}\bm{Q}'} V^s_{\bm{Q},\bm{Q}'}(\bm{k}-\bm{k}') \sum_{ll'\bm{G}\bm{G}'}\langle K,\bm{k}' |K, l, \bm{G}-\bm{Q}\rangle \langle K',-\bm{k}' |K', l', \bm{G}'+\bm{Q}'\rangle  
 \langle K',l', \bm{G}' | K', -\bm{k}\rangle  \langle K,l,\bm{G} | K, \bm{k}\rangle \\
& =  \sum_{\bm{Q}\bm{Q}'} V^s_{\bm{Q},\bm{Q}'}(\bm{k}-\bm{k}')  \sum_{l\bm{G} }\langle K,\bm{k}' |K, l, \bm{G}-\bm{Q}\rangle \langle K,l,\bm{G} | K, \bm{k}\rangle \sum_{l'\bm{G}'}\langle K, \bm{k} |K,l',\bm{G}' \rangle  \langle K, l', \bm{G}'-\bm{Q}' | K, \bm{k}' \rangle    \\
& =  \sum_{\bm{Q}\bm{Q}'} V^s_{\bm{Q},\bm{Q}'}(\bm{k}-\bm{k}')  \langle K,\bm{k}' |S(-\bm{Q})| K, \bm{k}\rangle 
\langle K, \bm{k} |S(\bm{Q}')| K, \bm{k}' \rangle    \\
& =  \sum_{\bm{Q}\bm{Q}'} V^s_{\bm{Q},\bm{Q}'}(\bm{k}-\bm{k}')  \langle K, \bm{k} |S(\bm{Q})| K, \bm{k}' \rangle^*
\langle K, \bm{k} |S(\bm{Q}')| K, \bm{k}' \rangle ,
\end{aligned},
\label{eq:bandinter1}
\end{equation}
where $|\tau, \bm k\rangle$ is a short-hand notation of $|\tau, m=1, \bm k\rangle$, the TRS relation $\langle K',l', \bm{G}' | K', -\bm{k}\rangle = \langle K, \bm{k} | K,l',-\bm{G}' \rangle$ is utilized, and the structure matrix is defined as $S(\bm{Q}) = \sum_{l\bm{G}} |K, l, \bm{G}+\bm{Q}\rangle \langle K,l,\bm{G}| $. We note that $S(0)$ is an identity matrix, with its dimension determined by the cutoff in the mori\'e reciprocal lattice vectors.  It can be straightforwardly shown that
\begin{equation}
\langle K, \bm{k} |S(\bm{Q})| K, \bm{k}' \rangle^*
\langle K, \bm{k} |S(\bm{Q}')| K, \bm{k}' \rangle = \sum_{ll'\bm{G}\bm{G}'} [\rho^{K}_{\bm{k}}]_{l,\bm{G};l',\bm{G}'} [\rho^{K}_{\bm{k'}}]_{l',\bm{G}'-\bm{Q}';l,\bm{G}-\bm{Q}} = F_{\bm{Q}',\bm{Q}}(\bm{k}',\bm{k}),
\label{eq:formfactorfmb}
\end{equation}
where $\rho_{\bm k}^{\tau}= |\tau, \bm k\rangle \langle \tau, \bm k|$ is the density matrix projected to the first moir\'e band.
Therefore, the pairing interaction defined in Eq.~(\ref{eq:bandinter1}) can be expressed as 
\begin{equation}
V_b( \bm{k}',\bm{k}) = \sum_{\bm{Q}\bm{Q}'} V^s_{\bm{Q},\bm{Q}'}(\bm{k}-\bm{k}') F_{\bm{Q}',\bm{Q}}(\bm{k}',\bm{k}).
\label{eq:bandint2}
\end{equation}
As illustrated diagrammatically in Fig.~2(b) of the main text, the linearized gap equation is derived as
\begin{equation}
\Delta_{\bm{k}} = -\frac{1}{A}\sum_{\bm{k}'}V_{b}(\bm{k},\bm{k}') \frac{\tanh(\beta \epsilon^{K}_{\bm{k}'}/2)}{2\epsilon^{K}_{\bm{k}'}}  \Delta_{\bm{k}'}
\label{eq:lgepband}
\end{equation}
where $\Delta_{\bm{k}}$ denotes the superconducting order parameter, $\beta = 1/k_BT$, $\epsilon^{K}_{\bm{k}'}$ is the band energy of the first mori\'e band in the $K$-valley. We note that, in this work, the linearized gap equation is solved within the first mori\'e band. This is because the energy scale of superconducting order (on the orders of hundreds of mK) is significantly smaller than the mori\'e band width (approximately 80 meV). Therefore, the inter-band pairing effects are negligible. 

The pairing interaction $V_{b}(\bm{k},\bm{k}')$ is a real valued function, as can be seen from  Eq.~\eqref{eq:bandinter1} and the first line in Eq.~\eqref{eq:relations}. As a result, the matrix appearing in the linearized gap equation, Eq.~\eqref{eq:lgepband}, contains only real entries. For real eigenvalues, such a matrix admits real eigenvectors, implying that the superconducting order parameter $\Delta_{\bm k}$ at $T_c$ can always be chosen to be real. In cases where two eigenvalues are degenerate due to rotational symmetry, the corresponding real eigenvectors, labeled $\Delta_1$ and $\Delta_2$, can be combined to form chiral pairing order parameters of the form $\Delta_1 \pm i \Delta_2$.
Another general feature of the order parameter $\Delta_{\bm k}$ is its periodicity in the MBZ, $\Delta_{\bm k + \bm G} = \Delta_{\bm k}$, where $\bm G$ is a moir\'e reciprocal lattice vector. This follows directly from the periodicity of the interaction, $V_{b}(\bm{k} + \bm {G},\bm{k}') = V_{b}(\bm{k},\bm{k}')$, along with the structure of the linearized gap equation.
These two property are general for pairing between two bands that are time-reversal partners, independent of the pairing mechanism.

\subsection{Single-band screening}
In this section, we focus on the screening effects arising from the first mori\'e band. Accordingly, the charge polarization function described in Eq.~(\ref{eq: chargeP}) simplifies to 
\begin{equation}
\begin{aligned}
\Pi_{\bm{Q},\bm{Q}'}(\bm{q}) &=  
\frac{2}{A}\sum_{\bm{k}}  \frac{f(\epsilon^{K}_{ \bm{k}})-f(\epsilon^{K}_{ \bm{k}+\bm{q}})}{\epsilon^{K}_{\bm{k}}-\epsilon^{K}_{ \bm{k}+\bm{q}}} 
F_{\bm{Q},\bm{Q}'} (\bm{k},\bm{k}+\bm{q}),
\end{aligned}
\end{equation}
where the form factor $F_{\bm{Q},\bm{Q}'} (\bm{k},\bm{k}+\bm{q})$ is defined in Eq.~(\ref{eq:formfactorfmb}).
By substituting the calculated $\Pi(\bm{q}) $ into Eq.~(\ref{eq: screenedCP}), and then inserting the resulting $V^s(\bm{q})$ into Eq.~(\ref{eq:bandint2}), we obtain the single-band screened pairing interaction $V_b(\bm{k},\bm{k}')$. The superconducting critical temperature ($T_c$) is determined by the condition that the largest eigenvalue of Eq.~(\ref{eq:lgepband}) equals 1.

\begin{figure*}
 \centering
\includegraphics[width=0.99\columnwidth]{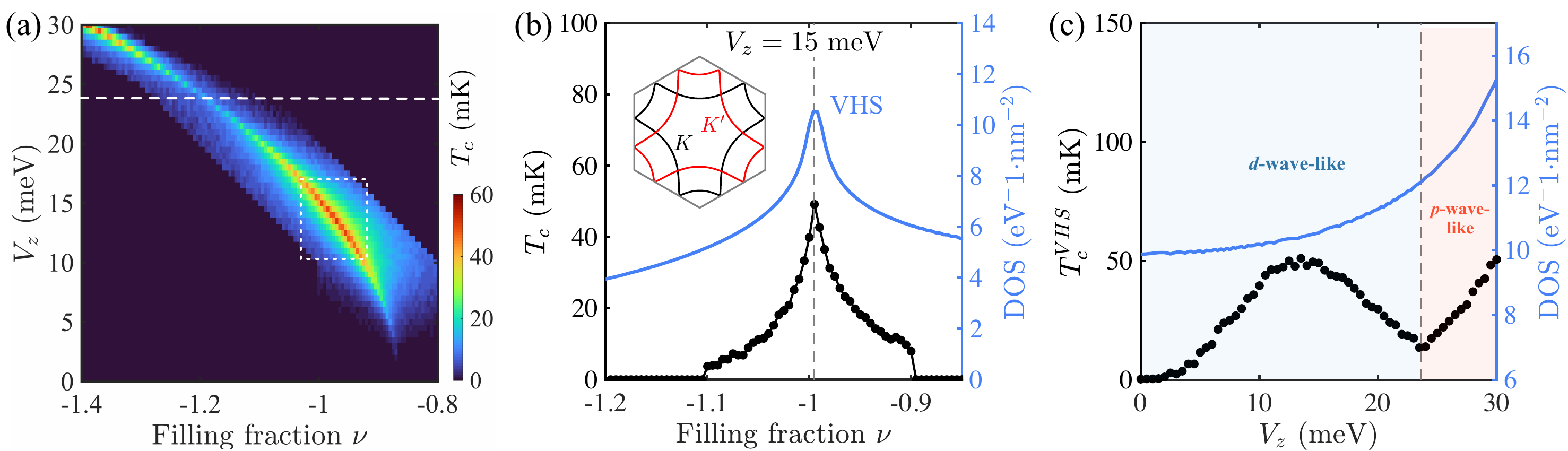}
 \caption{ \textbf{Single-band RPA:} (a) A two-dimensional map of the superconducting critical temperature $T_c$ versus filling fraction $\nu$ and layer potential $V_z$. The dotted rectangular highlights the $T_c$ maximum occurs at the VHS around $\nu \sim -1$, while the horizontal dashed line indicates the onset of a reentrant superconducting state.
 (b) $T_c$ (black dots) and DOS (blue curve) versus $\nu$ for $V_z =15$ meV ,where the inset depicts the $K$-valley (black) and $K'$-valley (red) FSs at the corresponding van Hove filling. (c) $T_c^{VHS}$ (black dots) and DOS (blue curve) at the VHSs versus $V_z$. The vertical dashed line indicates the onset of the reentrant superconducting state.}
 \label{fig:figureS1}
\end{figure*}

As shown in Fig.~\ref{fig:figureS1}, the single-band screened Coulomb interaction can induce pairing instabilities, with maximum $T_c$ on the order of tens of mK. 
Figure~\ref{fig:figureS1}(a) depicts the $T_c$ phase diagram as a function of the filling fraction $\nu$ and layer potential $V_z$. By comparing Figs.~\ref{fig:figureS1}(a) and Fig. 1(b) in the main text, we find the $T_c$ maxima consistently occur at the VHSs for given values of $V_z$. 
Figure~\ref{fig:figureS1}(b) shows the dependence of $T_c$ on $\nu$ for a fixed $V_z=15$ meV, wherein $T_c$ closely follows the DOS, showing a cusp at the VHS. These characters are indicative of superconductivity in the weak-coupling limit. 
In Fig.~\ref{fig:figureS1}(c), the values of $T_c^{VHS}$ and the corresponding DOS at VHSs are plotted for different values of $V_z$. 
Unlike the monotonically increasing DOS with $V_z$, $T_c^{VHS}$ displays a non-monotonic dependence. Specifically, as $V_z$ increases from 0 meV, $T_c^{VHS}$ initially rise to a maximum of $\sim$ 50 mK at $V_z = 14$ meV, subsequently decreases to a minimum at $V_z=23$ meV. The emergence of a minimum in $T_c$ at $V_z=23$ meV is associated with a first-order phase transition, as elaborated below.

\begin{figure}
 \centering
\includegraphics[width=0.99\columnwidth]{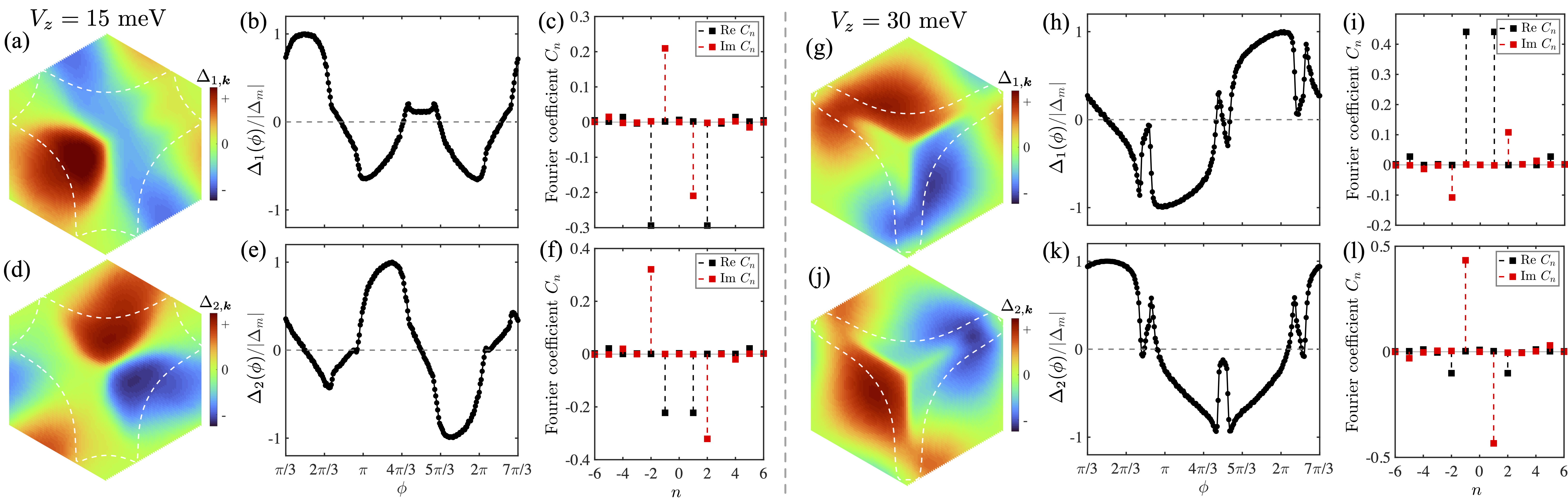}
 \caption{ \textbf{Order parameters:} (a)-(f) Superconducting order parameters for $V_z =15$ meV at the van Hove filling. (a),(d) $\bm{k}$-space structures of doubly-degenerate order parameters $\Delta_{1,\bm{k}}$ and $\Delta_{2,\bm{k}}$, where the dashed (white) curves denote the  corresponding FS. (b),(e) Samplings of $\Delta_{1,\bm{k}}$ and $\Delta_{2,\bm{k}}$ along the FS at equally spaced angles $\phi$ relative to the MBZ center (see the main text). The extracted angular profiles $\Delta_{1,2}(\phi)$ are normalized to their maximum values $|\Delta_{m}|$, respectively. (c),(f) Fourier decompositions of $\Delta_{1,2}(\phi)$ shown in (b) and (e), where $C_n$ denotes the Fourier coefficient of angular momentum channel $n$. (g)-(l) Superconducting order parameters for $V_z =30$ meV at the corresponding van Hove filling. (g),(j) $\bm{k}$-space structures of doubly degenerate order parameter $\Delta_{1,\bm{k}}$ and $\Delta_{2,\bm{k}}$. (h),(k) Samplings of $\Delta_{1,\bm{k}}$ and $\Delta_{2,\bm{k}}$ along the FS. (i),(l) Fourier 
 decompositions of $\Delta_{1,2}(\phi)$ shown in (h) and (k).}
 \label{fig:figureS21}
\end{figure}

We examine superconducting order parameters at the VHSs for two typical values of $V_z$. As shown in Fig.~\ref{fig:figureS21}(a), the leading pairing instability for $V_z =15$ meV is doubly degenerate, characterized by two orthogonal order parameters, $\Delta_{1,\bm{k}}$ and $\Delta_{2,\bm{k}}$, illustrated in Figs.~\ref{fig:figureS21}(a) and (d), respectively. 
The corresponding angular profiles $\Delta_{1,2}(\phi)$, extracted along FS, are shown in Figs.~\ref{fig:figureS21}(b) and (e), each exhibiting four nodes within a $2\pi$ period, indicating a $d$-wave-like pairing symmetry. Fourier decompositions of $\Delta_{1,2}(\phi)$, displayed in Figs.~\ref{fig:figureS21}(c) and (f), reveal dominant contributions to $\Delta_{1,2}(\phi)$ arise from $d$-wave channel with $n = \pm 2$.
For $V_z$ = 30 meV, the leading pairing instability is also doubly degenerate, with the $\bm{k}$-space structure of the order parameters shown in Figs.~\ref{fig:figureS21}(g) and (j). In contrast to the $V_z =15$ meV case, the dominant pairing symmetry is now $p$-wave-like, as confirmed by the Fourier decompositions in Figs.~\ref{fig:figureS21}(i) and (l), which exhibit leading components at $n = \pm 1$.

These quantitative analyses further confirm the presence of a first-order phase transition, characterized by a change in the dominant pairing symmetry as $V_z$ increases.

\subsection{Multi-band screening}
\label{subs:Mscreening}
\begin{figure*}
 \centering
\includegraphics[width=0.9\columnwidth]{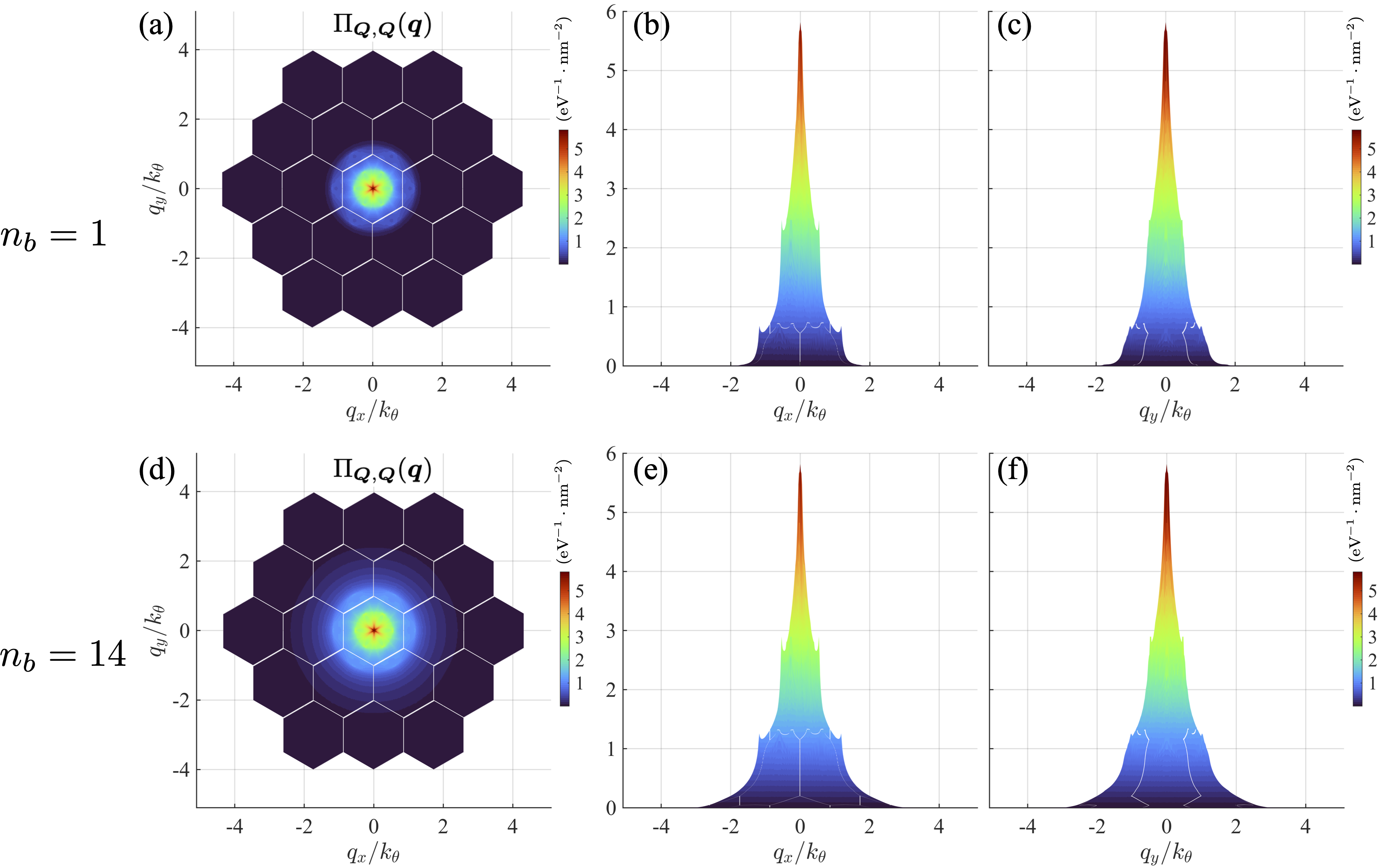}
 \caption{(a)-(c) Diagonal terms, $\Pi_{\bm{Q},\bm{Q}}(\bm{q})$, of the charge polarization function (as given in Eq.~(\ref{eq: chargeP})), calculated by including only the first mori\'e band, namely, $n_b =1$. (d)-(f) $\Pi_{\bm{Q},\bm{Q}}(\bm{q})$ calculated by choosing $n_b=14$. 
 Here Cartesian coordinate is adopted to display the data, where $(q_x, q_y) = \bm{q} +\bm{Q}$ with $\bm{q} \in$ MBZ. These results are obtained at the VHS for twist angle $\theta= 5^{\circ}$ and layer potential $V_z=15$ meV.  }
 \label{fig:figureS2}
\end{figure*}

\begin{figure*}
 \centering
\includegraphics[width=0.75\columnwidth]{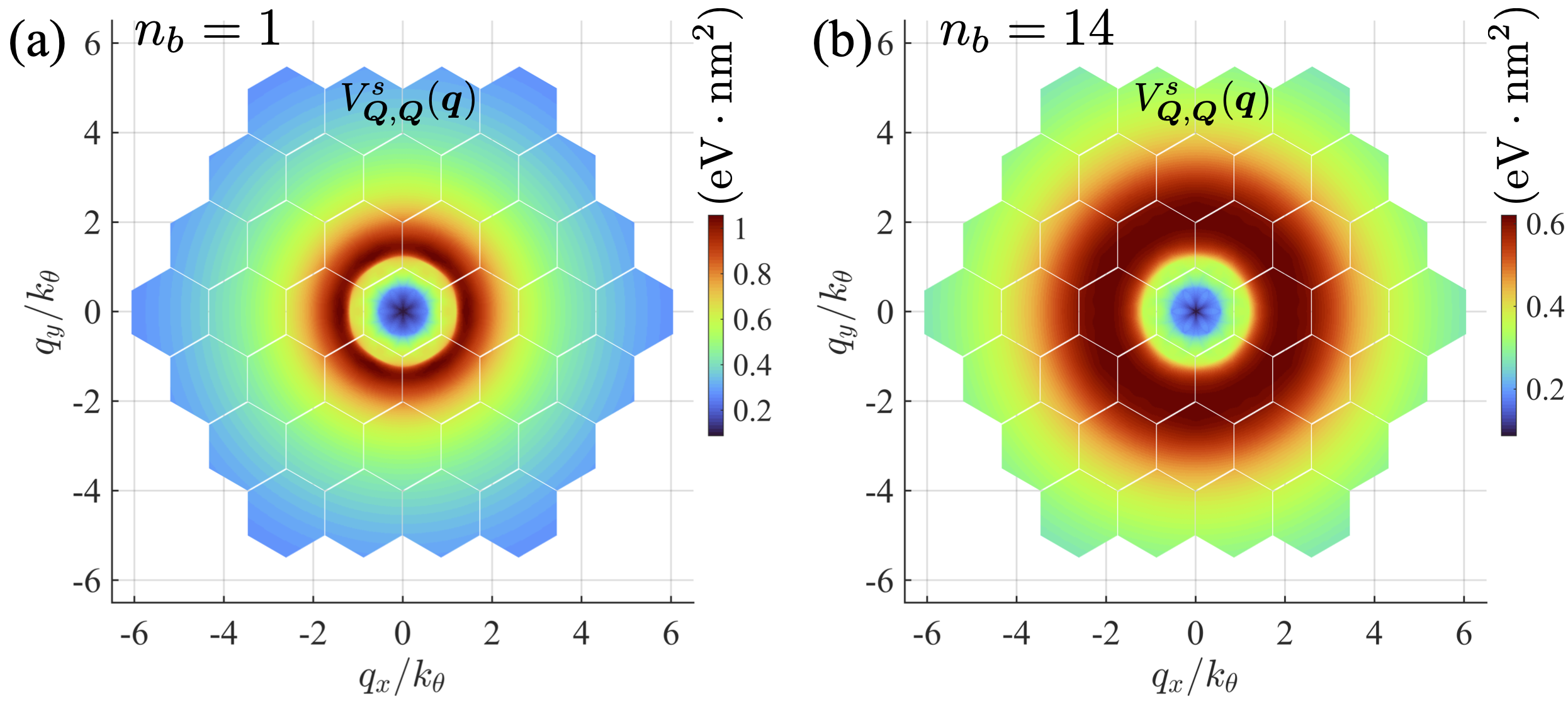}
 \caption{Diagonal terms, $V^s_{\bm{Q},\bm{Q}}(\bm{q})$, of the screened Coulomb potential (as given in Eq.~(\ref{eq: screenedCP})) for (a) $n_b =1$ and (b) $n_b=14$. Similar to Fig.~\ref{fig:figureS2}, the Cartesian coordinate is employed to display $V^s_{\bm{Q},\bm{Q}}(\bm{q})$.}
 \label{fig:figureS3}
\end{figure*}

In this section, we examine the impact of including multiple bands on the screening effect. The corresponding multiple-band charge polarization function is given by Eq.~(\ref{eq: chargeP}). Figure~\ref{fig:figureS2} compares the computed $\Pi_{\bm{Q},\bm{Q}'}(\bm{q})$ for two cases with different numbers of mor\'e bands: $n_b=1$ and $n_b= 14$. It is revealed that the dominant features of $\Pi_{\bm{Q},\bm{Q}'}(\bm{q})$ within the first MBZ is determined by the contributions from the first mori\'e band. The inclusion of additional remote bands leads to an enhancement  of $\Pi_{\bm{Q},\bm{Q}'}(\bm{q})$ away from the MBZ center and introduces additional contributions in the second and third reciprocal lattice shells.
Using the obtained $\Pi_{\bm{Q},\bm{Q}'}(\bm{q})$, we compute $V^s(\bm{q})$ based on Eq.~\eqref{eq: screenedCP}. As shown in Fig.~\ref{fig:figureS3}, the main features of the screened Coulomb potential within the first MBZ (central hexagon) are similar. Nevertheless, the Coulomb potential is suppressed for $n_b=14$, particularly around the edges of first MBZ. This behavior is reasonable, as the inclusion of multiple bands allows for additional inter-band particle-hole excitations, leading to a further screening of the Coulomb interaction.

\begin{figure}
 \centering
\includegraphics[width=0.75\columnwidth]{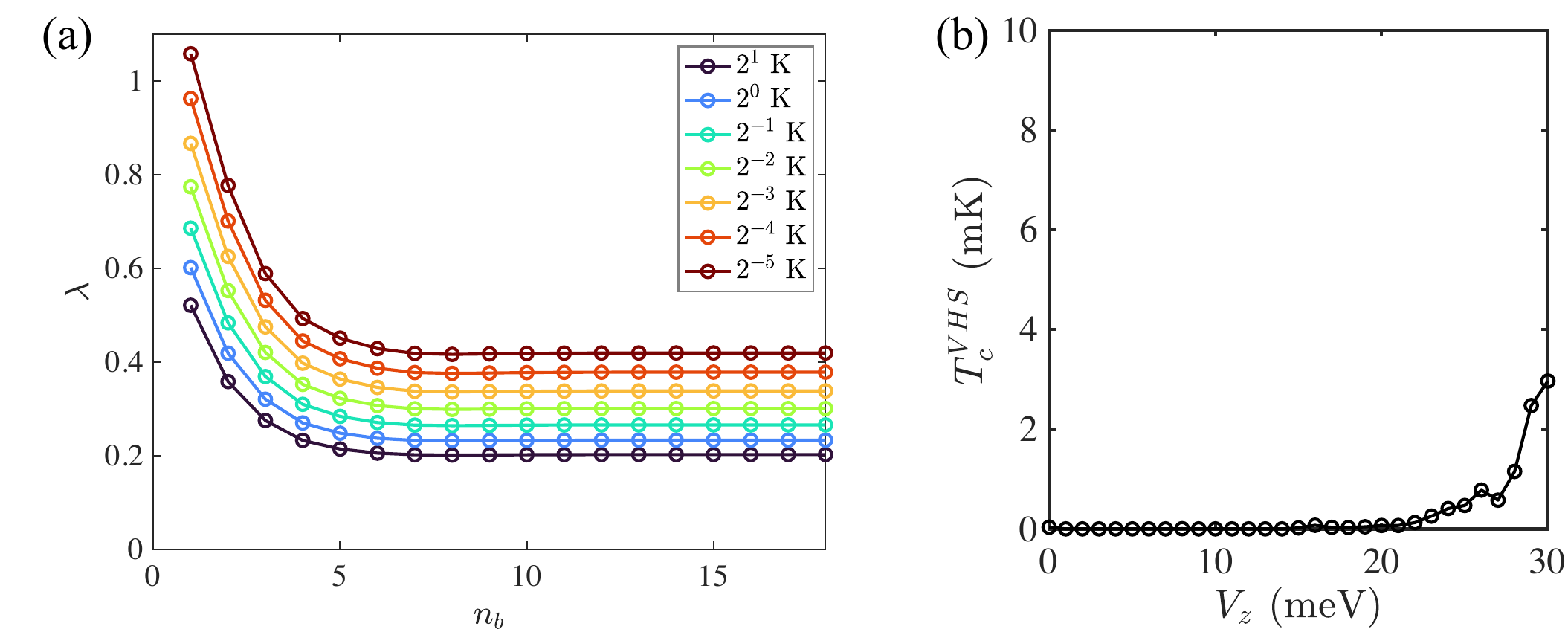}
 \caption{ \textbf{Multiple-bands RPA:} (a) Eigenvalues $\lambda$ of the linearized gap equation as function of the number of mori\'e bands and temperature. These results are obtained at the VHS for twist angle $\theta= 5^{\circ}$ and layer potential $V_z=15$ meV. (b) $T_c^{VHS}$ calculated at the VHSs versus $V_z$ for $n_b=14$.  }
 \label{fig:figureS5}
\end{figure}

Based on the obtained multiple-band screened Coulomb potential, we solve the linearized gap equation given by Eq.~(\ref{eq:lgepband}). Figure~\ref{fig:figureS5}(a) shows the largest eigenvalue ($\lambda$) of Eq.~(\ref{eq:lgepband}) as function of $n_b$ calculated for different temperatures. We find that $\lambda$ is quickly suppressed as $n_b$ increases, and it saturates when $n_b>8$. 
Figure~\ref{fig:figureS5}(b) shows the calculated $T_c's$ at the VHSs for different values of $V_z$. In these results, the screened Coulomb potential is computed by choosing $n_b=14$, where $\lambda$ is saturated. Comparing with the results shown Fig.~\ref{fig:figureS1}(c), we find that $T_c$ is dramatically suppressed to values that are beyond the detection limits of current experiments \cite{Guo:2024aa}. 

Based on these results, we conclude that the inclusion of remote bands in the RPA calculations results in stronger screening effect, which suppresses the pairing instability on the Fermi surface, resulting in undetectable superconductivity. Therefore, as discussed in the main text, the RPA-level Kohn-Luttinger mechanism cannot explain the experimentally observed superconductivity \cite{Guo:2024aa}.

\section{Inter-Valley Antiferromagnetic Fluctuations}
\label{sec:IVA_fluc}
\begin{figure}
 \centering
\includegraphics[width=0.9\columnwidth]{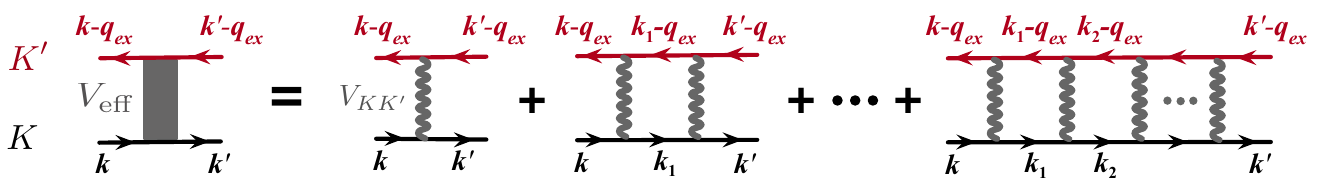}
 \caption{IVA fluctuation-induced effective interaction between electrons from opposite valleys. The filled rectangle represents the effective interaction $V_{\text{eff}}$, while the bold wavy lines denote the screened Coulomb interaction projected onto the first mori\'e band, labeled as $V_{KK'}$. }
 \label{fig:figureS6}
\end{figure}

Inspired by the experimental observation that superconductivity emerges near a metallic state with a FS reconstruction, which is believed to be associated with antiferromagnetic order, we explore the possibility of electron pairing mediated by inter-valley antiferromagnetic (IVA) fluctuations. Instead of starting with the bare Coulomb interaction, this work examines the IVA fluctuations on top of the multi-band RPA screened Coulomb potential, as derived in the preceding sections. Figure~\ref{fig:figureS6} shows the diagrammatic representation of the IVA fluctuations, where the building blocks of the ladder diagram are the inter-valley particle-hole susceptibility, $\chi_{KK'}(\bm{q}_{\text{ex}},\bm{k})$, and the RPA-screened Coulomb interaction. 
Due to the numerical computational challenges in treating the exchange diagrams, the present study considers the IVA fluctuations only from the first mori\'e band. The wavevector $\bm{k}$-resolved static inter-valley particle-hole susceptibility is given as
\begin{equation}
\chi_{KK'}(\bm{q}_{\text{ex}},\bm{k}) =  \frac{f(\epsilon^{K}_{ \bm{k}})-f(\epsilon^{K'}_{ \bm{k}-\bm{q}_{\text{ex}}})}{\epsilon^{K}_{\bm{k}}-\epsilon^{K'}_{ \bm{k}-\bm{q}_{\text{ex}}}}   = \frac{f(\epsilon^{K}_{ \bm{k}})-f(\epsilon^{K}_{ \bm{q}_{\text{ex}}-\bm{k}})}{\epsilon^{K}_{\bm{k}}-\epsilon^{K}_{ \bm{q}_{\text{ex}}-\bm{k}}},
\label{eq:k-resolvedPHS}
\end{equation}
where $\bm{q}_{\text{ex}}$ denotes the inter-valley nesting wavevector. In Fig.~\ref{fig:figureS6}, the IVA fluctuation-mediated effective interaction is formulated in the band representation. In this context, another key building block denoted by the bold wavy line is the band-projected screened Coulomb interaction. 
By projecting the mutiple-band screened Coulomb interaction obtained in Sec.~\ref{subs:Mscreening} to the first mori\'e band, we have the band-projected screened interaction
\begin{equation}
\begin{aligned}
V_{KK'}(\bm{k}',\bm{k}-\bm{q}_{\text{ex}};\bm{k}'-\bm{q}_{\text{ex}},\bm{k}) & = \sum_{\bm{Q}\bm{Q}'} V^s_{\bm{Q},\bm{Q}'}(\bm{k}-\bm{k}') \sum_{l'\bm{G}' } \langle K', \bm{k}-\bm{q}_{\text{ex}} |K',l',\bm{G}' +\bm{Q}' \rangle  \langle K', l', \bm{G}' | K', \bm{k}'-\bm{q}_{\text{ex}} \rangle  \\
&~~~~~~~~~~~~~~~~~~~~~~~~~~~~ \sum_{l\bm{G} }\langle K,\bm{k}' |K, l, \bm{G}-\bm{Q}\rangle \langle K,l,\bm{G} | K, \bm{k}\rangle   \\
& = \sum_{\bm{Q}\bm{Q}'} V^s_{\bm{Q},\bm{Q}'}(\bm{k}-\bm{k}') \sum_{l'\bm{G}'}\langle K, \bm{q}_{\text{ex}}-\bm{k}' |K,l',\bm{G}' +\bm{Q}'\rangle  \langle K, l', \bm{G}' | K, \bm{q}_{\text{ex}} - \bm{k} \rangle \\
&~~~~~~~~~~~~~~~~~~~~~~~~~~~~ \sum_{l\bm{G} }\langle K,\bm{k}' |K, l, \bm{G}-\bm{Q}\rangle \langle K,l,\bm{G} | K, \bm{k}\rangle    \\
&= \sum_{\bm{Q}\bm{Q}'} V^s_{\bm{Q},\bm{Q}'}(\bm{k}-\bm{k}') \langle K,  \bm{k} | S(\bm{Q}) | K, \bm{k}'\rangle^* \langle K,  \bm{q}_{\text{ex}} - \bm{k}' | S(\bm{Q}') | K,  \bm{q}_{\text{ex}} - \bm{k} \rangle 
\end{aligned}
\end{equation}
where the states denoted by $\bm{k}$ and $\bm{k}'$ belong to $K$ valley, while those labeled by $\bm{k}-\bm{q}_{\text{ex}}$ and $\bm{k}'-\bm{q}_{\text{ex}}$ belong to $K'$ valley. In the above derivation, the TRS is employed to simplify the expression. For a given $\bm{q}_{ex}$, $V_{KK'}(\bm{k}',\bm{k}-\bm{q}_{\text{ex}};\bm{k}'-\bm{q}_{\text{ex}},\bm{k})$ can be expressed as a matrix $V_{KK'}(\bm{q}_{\text{ex}})$ whose elements are indexed by the in-going ($\bm{k}$) and out-going ($\bm{k}'$) wavevectors. As illustrated in Fig.~\ref{fig:figureS6}, the IVA fluctuation-mediated effective interaction $V_{\text{eff}}(\bm{q}_{ex})$ can likewise be represented in a matrix form. Consequently, the ladder diagram summation shown in Fig.~\ref{fig:figureS6} can be compactly formulated as
\begin{equation}
V_{\text{eff}}(\bm{q}_{\text{ex}}) = \frac{V_{KK'}(\bm{q}_{\text{ex}})}{\mathbb{1}+\chi_{KK'}(\bm{q}_{\text{ex}}) V_{KK'} (\bm{q}_{\text{ex}}) },
\label{eq:IVAfluct}
\end{equation}
where $\chi_{KK'}(\bm{q}_{\text{ex}}) = \chi_{KK'}(\bm{q}_{\text{ex}},\bm{k}) \delta_{\bm{k},\bm{k}'}$ is represented as a diagonal matrix in the momentum space.
Compared to the RPA-level calculation of the screened Coulomb potential, the calculation of the IVA fluctuation-mediated interaction $V_{\text{eff}}$ is significantly more demanding. This is primarily due to the much larger matrix dimensions of $V_{KK'}(\bm{q}_{\text{ex}})$ and $V_{\text{eff}}(\bm{q}_{\text{ex}})$ compared to those of $\Pi(\bm{q})$ and $V^s(\bm{q})$ in Eq.~(\ref{eq: screenedCP}). To manage this computational complexity, in this work, we restrict our calculations of $V_{\text{eff}}(\bm{q}_{\text{ex}})$ to contributions from the first mori\'e band. We note that omitting contributions from other mori\'e bands likely underestimate the exchange enhancement of $V_{\text{eff}}(\bm{q}_{\text{ex}})$.

\subsection{Pairing instability}
In this section, we examine the possibility of the effective interaction-induced pairing instability between electrons from opposite valleys with opposite momenta. This corresponds to the zero-momentum pairing state, which requires the total in-going wavevector shown in Fig.~\ref{fig:figureS6} to be zero, leading to the condition $\bm{q}_{\text{ex}} = \bm{k}+\bm{k}'$. Under this constraint, the effective pairing interaction reduces to $V_{\text{eff}}(\bm{k}',-\bm{k}';-\bm{k},\bm{k})$, giving rise to the linearized gap equation for superconductivity presented as Eq.~(7) in the main text.

In analogy to the antiferromagnetic fluctuation-mediated superconductivity in high-$T_c$ superconductors, the IVA fluctuations here can enhance the pairing interaction, as suggested by Eq.~(\ref{eq:IVAfluct}), where the denominator is less than unity. As demonstrated in Fig.~3 in the main text, this enhancement results in substantially increased superconducting critical temperatures, reaching values on orders of hundreds of miliKelvins, comparable to experimental observations. Combined with the multiple-band screening results presented in Sec.~\ref{subs:Mscreening}, we conclude that IVA fluctuations serve as the primary pairing mechanism in the present system.

\subsection{IVA ordering}
The strength of IVA fluctuation is characterized by inter-valley particle-hole susceptibility $\chi_{KK'}(\bm{q}_{\text{ex}})$, which is negative by definition, as given in Eq.~(\ref{eq:k-resolvedPHS}). If the IVA fluctuations become sufficiently strong, the system tends to develop IVA-ordered states, which compete with superconductivity. Similar to the pairing instability, a linearized gap equation describing the onset of IVA order is diagrammatically illustrated in Fig.~2(c) in the main text and is expressed formally in Eq.~(8). 

\begin{figure}
 \centering
\includegraphics[width=0.9\columnwidth]{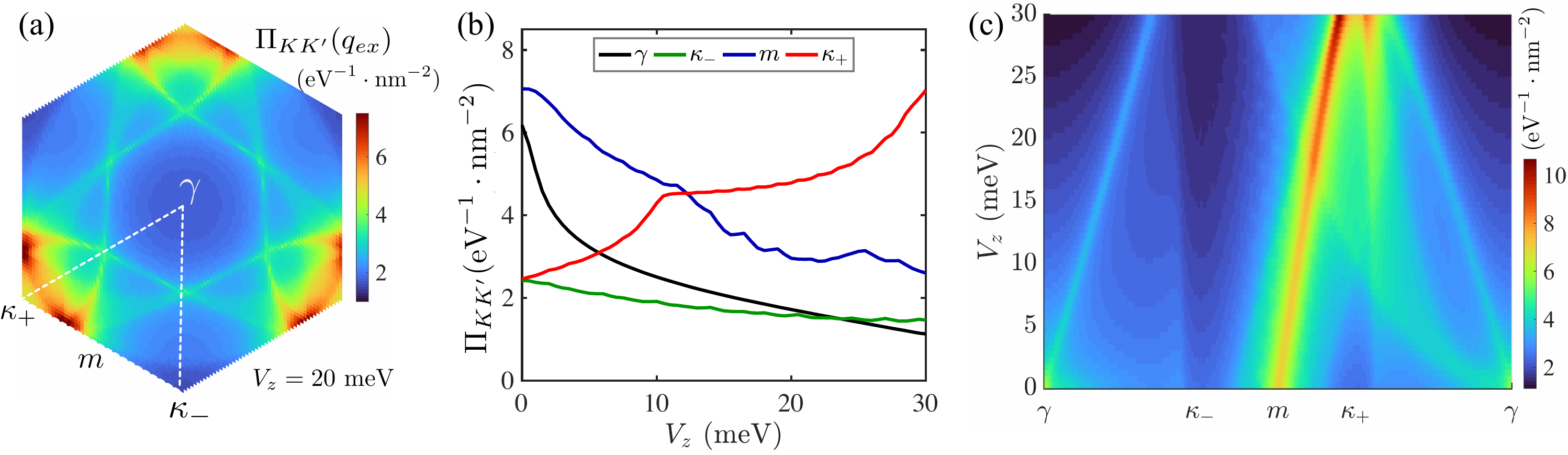}
 \caption{ \textbf{Inter-valley particle-hole susceptibility $\Pi_{KK'}(q_{\text{ex}})$:} (a) 2D color map of $\Pi_{KK'}(q_{\text{ex}})$, where nesting wavevector $q_{\text{ex}}$ varies over the MBZ, evaluated at a fixed layer potential $V_z =20$ meV. (b) $\Pi_{KK'}(q_{\text{ex}})$ as a function of $V_z$, with $q_{\text{ex}}$ fixed at high-symmetry points of MBZ: $\gamma$, $\kappa_{-}$, $m$, and $\kappa_{+}$. (c) 2D map of $\Pi_{KK'}(q_{\text{ex}})$ as a function of $V_z$ and $q_{\text{ex}}$ along the high-symmetry paths indicated by dashed lines in panel (a). 
 These results are calculated at the VHSs for various values of $V_z$ and at temeprature $T=50$ mK.}
 \label{fig:figureS7}
\end{figure}

Here we provide a perspective based on the inter-valley particle-hole susceptibility.
Although the computation of $V_{\text{eff}}(\bm{q}_{\text{ex}})$ in Eq.~(\ref{eq:IVAfluct}) essentially involves the $\bm{k}$-resolved particle-hole susceptibility $\chi_{KK'}(\bm{q}_{\text{ex}},\bm{k}) $, it is also instructive to examine the total inter-valley particle-hole susceptibility, which is defined as 
 \begin{equation}
 \Pi_{KK'}(\bm{q}_{\text{ex}}) = \frac{1}{A}\sum_{\bm{k}} \chi_{KK'}(\bm{q}_{\text{ex}},\bm{k}) =   \frac{1}{A}\sum_{\bm{k}} \frac{f(\epsilon^{K}_{ \bm{k}})-f(\epsilon^{K}_{ \bm{q}_{\text{ex}}-\bm{k}})}{\epsilon^{K}_{\bm{k}}-\epsilon^{K}_{ \bm{q}_{\text{ex}}-\bm{k}}}.
 \end{equation}
 Figure~\ref{fig:figureS7}(a) presents the calculated $ \Pi_{KK'}(\bm{q}_{\text{ex}}) $ at the VH filling for layer potential $V_z=20$ meV. Three pronounced peaks are observed surrounding the $\kappa_{+}$ point in the MBZ. These maxima are attributed to the enhanced inter-valley nesting between saddle points in the band structure.  
 
Figure~\ref{fig:figureS7}(b) shows the dependence of $\Pi_{KK'}(\bm{q}_{\text{ex}})$ on $V_z$ for four commensurate nesting wavevectors $\bm{q}_{ex} = \gamma$, $m$, and $\kappa_{\pm}$ in the MBZ. For $V_z=0$ meV, $\Pi_{KK'}(\bm{\gamma})$ reaches its maximum value, corresponding to the strongest channel of the IVA fluctuations. This behavior is attributed to the presence of saddle point at the $m$ point of the MBZ. 
For $V_z>12$ meV, $\Pi_{KK'}(\bm{\kappa_{+}})$ dominates over other commensurate IVA fluctuation channels due to the migration of the saddle point toward $\kappa_{+}$. As illustrated in Fig.~3(a) of the main text, the SC phase predominantly emerges for $V_z > 12 $ meV. This motivates our focus on the competing between superconductivity and the IVA order characterized by $\bm{q}_{ex}= \kappa_{+}$ throughout this study. 

Figure \ref{fig:figureS7}(c) depicts the dependence of $\Pi_{KK'}(\bm{q}_{\text{ex}})$ on $V_z$ along the high-symmetry paths of the MBZ. A prominent feature is the shift of the strongest IVA fluctuation from the $m$ point toward the $\kappa_{+}$ point upon increasing $V_z$. This trends closely follows the migration of the saddle point along the $m-\kappa_{+}$ direction. In principle, incommensurate IVA orders may also compete with SC. In this work, we focus primarily on the commensurate IVA orders for the reasons outlined in the following paragraph.

The maxima of $\Pi_{KK'}(\bm{q}_{\text{ex}})$ occurs at three incommensurate wavevectors on the edges of the MBZ, as illustrated in Fig.~\ref{fig:figureS7}(a). These wavevectors lie in close proximity to the $\kappa_{+}$ point within in the parameter regime ($V_z>12$ meV) where pronounced superconductivity is observed. The presence of multiple prominent nesting wavevectors around $\kappa_{+}$ point may lead to competition between different incommensurate IVA orders. Such competition can enhance fluctuations that ultimately favor the formation of a commensurate IVA order at wavevector $\kappa_{+}$ \cite{Munoz-Segovia:2025aa}. This commensurate order would be further stabilized at low temperatures or under strong interaction strength, due to the additional energy gain from umklapp processes--often referred to "lock-in" energy \cite{McMillan:1975aa,Pokrovsky:1979aa}.

\section{Self-consistent mean-filed calculation}
In the preceding sections, we solved the linearized gap equation near the superconducting transition, which yields both the critical temperature and the pairing symmetry. In this section, we perform self-consistent mean-field calculations of the superconducting state below $T_c$ to further examine the stability of the superconducting phase. The Hamiltonian including pairing interaction is derived as 
\begin{equation}
H = \sum_{\tau \bm{k}} \epsilon_{\bm{k}}^{\tau } c_{\tau,\bm{k}}^{\dagger} c_{\tau, \bm{k}}  + \frac{1}{A} \sum_{\bm{k}\bm{k}'} V_{\text{eff}}(\bm{k}',-\bm{k}';-\bm{k},\bm{k}) c_{K,\bm{k}'}^{\dagger} c_{K',-\bm{k}'}^{\dagger} c_{K',-\bm{k}} c_{K,\bm{k}}.
\end{equation}
By performing mean-field decoupling to the interaction Hamiltonian (the second term on the right-hand side of the above equation), we obtain the mean-filed Hamiltonian
\begin{equation}
H_{\text{MF}} = \sum_{\bm{k}} \Psi_{\bm{k}}^{\dagger} \mathcal{H}_{\text{BdG}}(\bm{k}) \Psi_{\bm{k}} + \sum_{\bm{k}} \epsilon_{\bm{k}}^{K} -\sum_{\bm{k}} \Delta_{\bm{k}} \langle c_{K,\bm{k}}^{\dagger} c_{K',-\bm{k}}^{\dagger} \rangle,
\end{equation}
where $ \Psi_{\bm{k}} = (c_{K,\bm{k}}, c_{K',-\bm{k}}^{\dagger})^{\text{T}}$, the pairing potential $\Delta_{\bm{k}'}=\frac{1}{A} \sum_{\bm{k}\bm{k}'} V_{\text{eff}}(\bm{k}',-\bm{k}';-\bm{k},\bm{k}) \langle c_{K',-\bm{k}} c_{K,\bm{k}} \rangle$, and the Bogoliubov–de Gennes (BdG) Hamiltonian
\begin{equation}
\mathcal{H}_{\text{BdG}}(\bm{k})  = 
\begin{pmatrix}
\epsilon_{\bm{k}}^{K} & \Delta_{\bm{k}}  \\
 \Delta_{\bm{k}}^{\dagger}  & -\epsilon_{\bm{k}}^{K}
\end{pmatrix}.
\label{eq:BdG}
\end{equation}
In the above derivation, the TRS relation $\epsilon_{\bm{k}}^{K} = \epsilon_{-\bm{k}}^{K'}$ has been utilized. 
The self-consistent gap equation is given by
\begin{equation}\label{eqt27}
\Delta_{\bm{k}'} = -\frac{1}{2A}\sum_{\bm{k}}V_{\text{eff}}(\bm{k}',-\bm{k}';-\bm{k},\bm{k}) \frac{\tanh(\beta E^{K}_{\bm{k}}/2)}{E^{K}_{\bm{k}}}  \Delta_{\bm{k}},
\end{equation}
where $E^{K}_{\bm{k}} =\sqrt{(\epsilon_{\bm{k}}^{K})^2+\Delta_{\bm{k}}^2} $ denotes the excitation spectrum of BdG quasiparticle, which is obtained by diagonalizing Eq.~(\ref{eq:BdG}). The free energy of the superconducting state is defined as $F_s=\Omega + N\mu$, where $\Omega = -\frac{1}{\beta} \ln \text{Tr} [e^{-\beta H_{\text{MF}}}] $ is the thermal dynamic grand potential, $N$ and $\mu $ denote the conserved particle number and the chemical potential, respectively. By inserting the mean-field Hamiltonian into the definition of $\Omega$, we have
\begin{equation}
\label{eq:free_energy}
F_s = -\frac{2}{\beta} \sum_{\bm{k}} \ln (1+e^{-\beta E_{\bm{k}}^{K}} )   +  \sum_{\bm{k}} (\epsilon_{\bm{k}}^{K}-E^{K}_{\bm{k}}) -\sum_{\bm{k}} \Delta_{\bm{k}} \langle c_{K,\bm{k}}^{\dagger} c_{K',-\bm{k}}^{\dagger} \rangle +N \mu.
\end{equation}
The superconducting condensation energy is derived as 
\begin{equation}
\label{eq:condensation_energy}
E_c = F_s-F_n,
\end{equation}
where $F_n$ represents the normal-state free energy, evaluated by setting $\Delta_{\bm{k}} = 0$ in Eq.~(\ref{eq:free_energy}). In principle, the chemical potential $\mu$ should also be determined self-consistently in the superconducting state by enforcing particle number conservation. However, in the present study, the self-consistent solutions of $\Delta_{\bm{k}} $ are on the order of a few $\mu$eV (see Fig.~5 in the main text), and therefore have a negligible effect on the renormalization of $\mu$ compared to its value in the normal state. As a result, the condensation energy $E_c$ is computed by keeping $\mu$ fixed in both the normal and superconducting states.

Since the superconducting order parameter is momentum-dependent, an appropriate choice of initial seed functions in the self-consistent calculations is crucial for efficiently converging to the superconducting ground state. In this study, we use the eigenvectors obtained from solving the linearized gap equation, scaled by an infinitesimally small factor, as the initial seed function. This approach results in rapid convergence to the superconducting ground state.

\section{Numerical Calculations}
\begin{figure}
 \centering
\includegraphics[width=0.5\columnwidth]{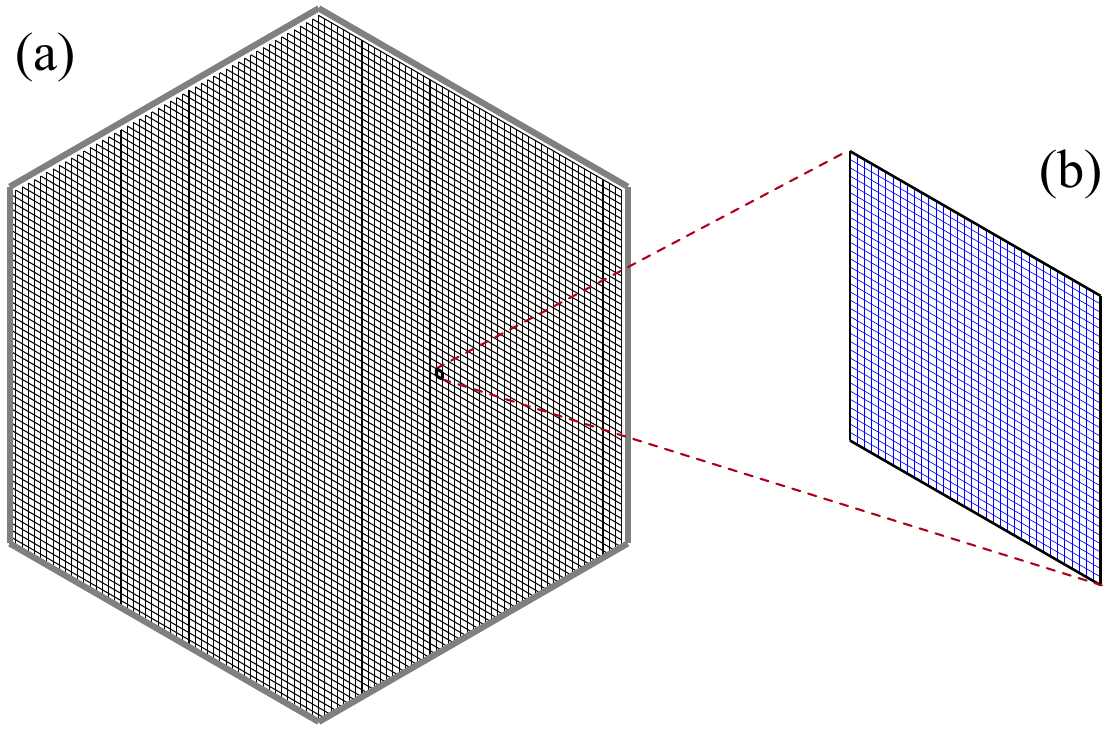}
 \caption{\textbf{Two-step discretization of the MBZ:} (a) The MBZ is divided into $N_p$ primary meshes to provide a smooth description of the wave function effect. (b) Each primary mesh is further divided into $N_s$ submeshes to give a more precise description of the Fermi surface.}
 \label{fig:figureS8}
\end{figure}

In this section, we provide numerical details on the discretization of the MBZ used in the calculations presented in the preceding sections and the main text. As indicated by the right-hand side of Eq.~(\ref{eq:ChargPF1}), the charge polarization function $\Pi_{\bm{Q},\bm{Q}'}^{\tau \tau}(\bm{q})$ is determined by two key factors: the band energy effect and wave function effect. They are characterized by $[f(\epsilon^{\tau}_{m \bm{k}})-f(\epsilon^{\tau}_{m' \bm{k}+\bm{q}})]/(\epsilon^{\tau}_{m \bm{k}}-\epsilon^{\tau}_{m' \bm{k}+\bm{q}})$ and form factor $F^{\tau m,\tau m'}_{\bm{Q},\bm{Q}'} (\bm{k},\bm{k}+\bm{q})$, respectively. The former is highly sensitive to the Fermi surface due to the presence of Fermi-Dirac distribution function, while the later remains relatively robust. To balance computational accuracy and efficiency, we adopt a two-step discretization of the MBZ, as schematically illustrated in Fig.~\ref{fig:figureS8}. In the first step, the MBZ is divided into $N_{p}$ primary meshes to accurately capture the form factor-related contributions. In the second step, each primary mesh is further refined into $N_s$ submeshes, as illustrated in Fig.~\ref{fig:figureS8}(b), enabling a more precise treatment of the band effect. Under this two-step discretization, Eq.~(\ref{eq:ChargPF1}) can be expressed as 
\begin{equation}
\begin{aligned}
\Pi_{\bm{Q},\bm{Q}'}^{\tau \tau}(\bm{q}) & = \frac{1}{ A} \sum_{ mm' } \sum_{\bm{k}\in N_p}F^{\tau m,\tau m'}_{\bm{Q},\bm{Q}'} (\bm{k},\bm{k}+\bm{q}) \left[\frac{1}{N_s}\sum_{\delta \bm{k} \in N_s} \frac{f(\epsilon^{\tau}_{m \bm{k}+\delta \bm{k}})-f(\epsilon^{\tau}_{m' \bm{k}+\delta \bm{k}+\bm{q}})}{\epsilon^{\tau}_{m \bm{k}+\delta \bm{k}}-\epsilon^{\tau}_{m' \bm{k}+\delta \bm{k}+\bm{q}}}   \right],
\end{aligned}
\end{equation}
where $\bm{k}$ denotes the center-of-mesh wavevector shown in Fig.~\ref{fig:figureS8}(a), and $\delta \bm{k}$ is measured from $\bm{k}$, representing the wavevector of the submeshes depicted in Fig.~\ref{fig:figureS8}(b). In the RPA calculations shown in Sec.~\ref{sec:RPA_screening}, we have adopted $N_p = 7500$ and $N_s=225$, which are sufficient to give a precise description of the wave function and band effects.

This two-step discretization scheme is also employed in solving the linearized gap equation. As shown in Eq.~(\ref{eq:lgepband}), 
the pairing interaction $V_b(\bm{k},\bm{k}')$ varies relatively smoothly across the MBZ and can be adequately captured using the primary mesh. In contrast, the particle-particle channel susceptibility, given by $\tanh(\beta \epsilon^{K}_{\bm{k}'}/2)/2\epsilon^{K}_{\bm{k}'}$, typically exhibits a logarithmic divergence at low temperatures. Precisely capturing this behavior requires a much denser $\bm{k}$ mesh. By applying the same two-step discretization to the MBZ, the linearized gap equation in Eq.~(\ref{eq:lgepband}) can be reformulated as
\begin{equation}
\Delta_{\bm{k}} = -\frac{1}{A}\sum_{\bm{k}' \in N_p}V_{b} (\bm{k},\bm{k}')  \Delta_{\bm{k}'} \left[\frac{1}{N_s} \sum_{\delta \bm{k}' \in N_s} \frac{\tanh(\beta \epsilon^{K}_{\bm{k}'+\delta \bm{k}'}/2)}{2\epsilon^{K}_{\bm{k}'+\delta \bm{k}'}} \right]  .
\end{equation}
As demonstrated in Figs.~\ref{fig:figureS1} and \ref{fig:figureS5}, this two-step discretization scheme yields smooth variations of the superconducting critical temperature $T_c$ to within a few mK upon changing the filling factors and layer potential.

In Sec.~\ref{sec:IVA_fluc}, we adopted $N_p = 1875$ and $N_s = 10^4$ for calculating the IVA fluctuation-mediated effective interaction, as these calculations are significantly more computationally demanding. Nevertheless, this primary mesh resolution provides a sufficiently smooth description of the resulting superconducting order parameters, as shown in Fig.~4 in the main text. 

As explained in the main text, the saddle point is located at the $m$ point of the MBZ for $V_z=0$ meV and shifts toward the $\kappa_{+}$ point along the MBZ edge as $V_z$ increases. To accurately capture the position of the saddle point using $N_p = 1875$ primary meshes, precise control of $V_z$ is required to ensure that the incremental shift of the saddle point remains compatible with the mesh resolution. 
As a result, only a finite set of $V_z$ values allows the saddle point to coincide with one of the mesh points, and the phase diagram shown in Fig.~3 of the main text is constructed using this discrete set of $V_z$ values. We note that, although $V_z$ is sampled discretely, this does not affect the qualitative features of the phase diagram, particularly the interplay between superconductivity and IVA fluctuations.

\section{inter-valley pairing favored by displacement field}\label{ipfbd}
\begin{figure}
 \centering
\includegraphics[width=0.75\columnwidth]{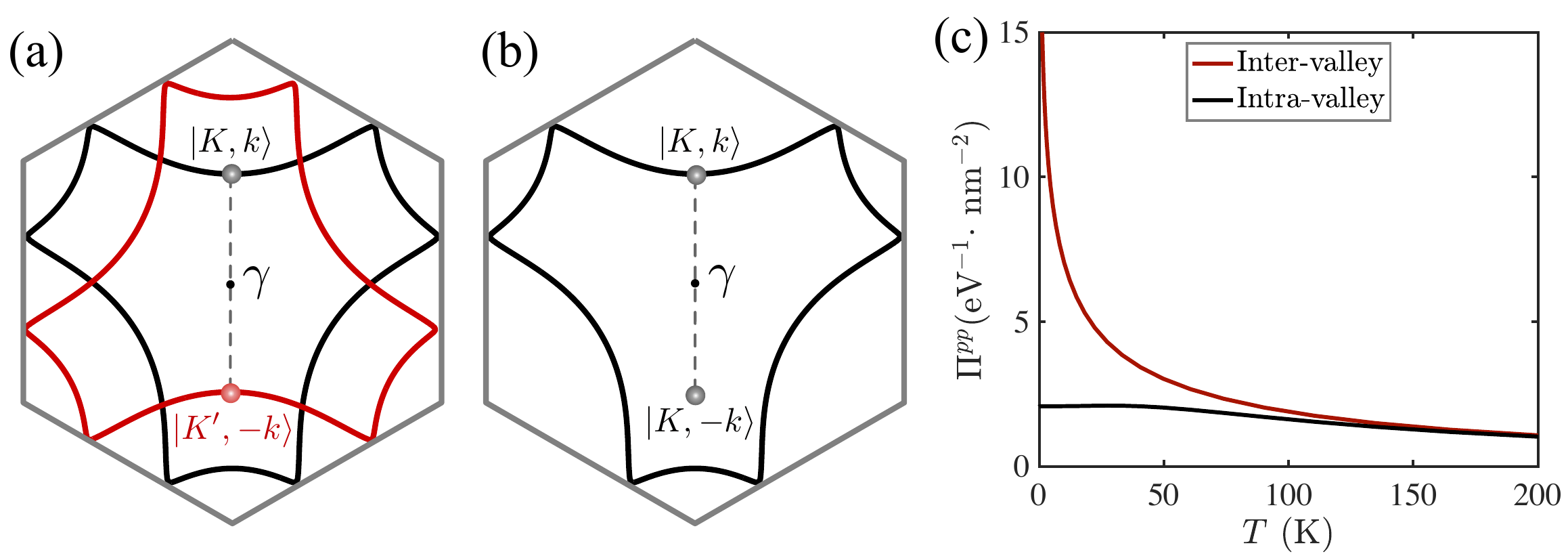}
 \caption{(a),(b) Schematic illustrations of electron pairing in tWSe$_2$: (a) intervalley pairing between electron from opposite valleys and (b) intra-valley pairing within the same valley. The black and red curves denote the $K$- and $K'$-valley Fermi surfaces, respectively. (c) Temperature dependence of the inter-valley (red) and intra-valley (black) particle-particle susceptibilities. These results are obtained at the van Hove filling corresponding to layer potential $V_z=15$ meV.}
 \label{fig:figureS9}
\end{figure}

In the presence of displacement field, the typical FS of tWSe$_2$ are illustrated in Fig.~\ref{fig:figureS9}, where the valley-resolved FS lacks inversion symmetry with respect to the center of MBZ. This asymmetry arises from the displacement field-induced symmetry lowering effect, which reduces the point group symmetry of the system to $C_3$. In this work, we focus on zero center-of-mass momentum pairing relative to the MBZ center. Figures~\ref{fig:figureS9}(a) and (b) schematically depict the inter-valley and intra-valley paring configurations. In the case of inter-valley pairing, as shown in Fig.~\ref{fig:figureS9}(a), if a state $|K,\bm k\rangle$ lies on the $K$-valley FS, its paired partner $|K',-\bm k\rangle$ lies on the $K'$-valley FS, as guaranteed by the time-reversal symmetry. In contrast, for the intra-valley pairing, as shown in Fig.~\ref{fig:figureS9}(b), its paired partner $|K,-\bm k\rangle$ generally does not lie on the $K$-valley FS due to the lack of inversion symmetry, thereby making the intra-valley pairing energetically unfavorable.

This difference can be quantitatively reflected in the inter-valley and intra-valley particle-particle susceptibilities, which quantify the tendency toward the pairing instability. Specifically, the inter-valley particle-particle susceptibility is defined as 
\begin{equation}
\Pi_{K K^{\prime}}^{p p}=\frac{1}{A} \sum_{\bm k} \frac{2 f\left(\epsilon_{\boldsymbol{k}}^K\right)-1}{2 \epsilon_{\boldsymbol{k}}^K},
\end{equation}
where $A$ is the system area, and $\epsilon_{\boldsymbol{k}}^K$ denotes the $K$-valley electron dispersion of the first moir\'e band. Similarly, the intra-valley particle-particle susceptibility is given by
\begin{equation}
\Pi_{K K}^{p p}=\frac{1}{A} \sum_{\bm k} \frac{f\left(\epsilon_{\bm k}^K\right)+f\left(\epsilon_{-{\bm k}}^K\right)-1}{\epsilon_{\bm k}^K+\epsilon_{-{\bm k}}^K}.
\end{equation}
Fig.~\ref{fig:figureS9}(c) plots the numerical results, where $\Pi_{K K^{\prime}}^{p p}$ exhibits a divergent behavior upon lowering temperature, while $\Pi_{K K}^{p p}$ saturates to a finite value at low temperatures. This behavior clearly indicates that the inter-valley pairing instability dominates over the intra-valley pairing.

\section{structure of order parameter}
The WSe$_2$ exhibits strong Ising-type spin-orbit coupling, which gives rise to a pronounced spin-valley locking effect: electrons in the $K$ and $K'$ valleys have their spins locked to opposite out-of-plane directions. As demonstrated in Sec.~\ref{ipfbd}, electron pairing is more favorable between opposite valleys. Therefore, the superconducting order parameter can be expressed as
\begin{equation}
\Delta(\boldsymbol{k})=i\left[\phi(\boldsymbol{k})+d(\boldsymbol{k}) \tau_z\right] \tau_y=\left[\begin{array}{cc}
0 & \phi(\boldsymbol{k})+d(\boldsymbol{k}) \\
-\phi(\boldsymbol{k})+d(\boldsymbol{k}) & 0
\end{array}\right],
\end{equation}
where the Nambu spinor is $ \Psi_{\bm{k}} = (c_{K,\bm{k}}, c_{K',\bm{k}}, c^{\dagger}_{K,-\bm{k}}, c_{K',-\bm{k}}^{\dagger})^{\text{T}}$, the Pauli matrices $\tau_{z,y}$ acting in the valley subspace (or equivalently, the spin subspace due to spin-valley locking). The function $\phi(\boldsymbol{k})=\phi(-\boldsymbol{k})$ represents the spin-singlet (valley-singlet) component, while $d(\boldsymbol{k})=-d(-\boldsymbol{k})$ denotes the spin-triplet (valley-triplet) component. In the absence of displacement field, the system exhibits an emergent inversion symmetry. For this case, the even-parity ($\phi(\boldsymbol{k})$, spin-singlet) and odd-parity ($d(\boldsymbol{k})$, spin-triplet) components of order parameter are decoupled. The presence of a finite displacement field breaks inversion symmetry, leading to the mixing of even- and odd-parity components, i.e., singlet-triplet mixing.

In our numerical calculations, we do not pre-decompose the order parameter into singlet and triplet components. Nevertheless, our computation approach is fully general and unbiased, without any symmetry constraints imposed on the structure of the order parameter. It therefore retains the full capability to capture both the even- and odd-parity contributions, which are mixed in the presence of the displacement field. As shown in Fig.~4 of the main text, the leading pairing instability is doubly degenerate and transforms under the $E$ irreducible representation of the $C_3$ point group. As expected, the order parameter contains both even- and odd-parity components. This is further confirmed by Fourier decomposition of the Fermi surface resolved order parameters, as shown in Figs.~4(e) and (f) in the main text, which show contributions from multiple angular momentum channels, including the $p$-wave (odd), $d$-wave (even), and higher harmonics. These results clearly indicate that the superconducting pairing state in our study comprises spin singlet-triplet mixing due to the broken of inversion symmetry.

Although our method does not involve a prior singlet-triplet decomposition, it provides an unbiased solution of the order parameter in the full-$\bm{k}$ space, and is thus fully capable of capturing the correct pairing state.

\section{Dielectric constant-dependence}
\begin{figure}
 \centering
\includegraphics[width=0.5\columnwidth]{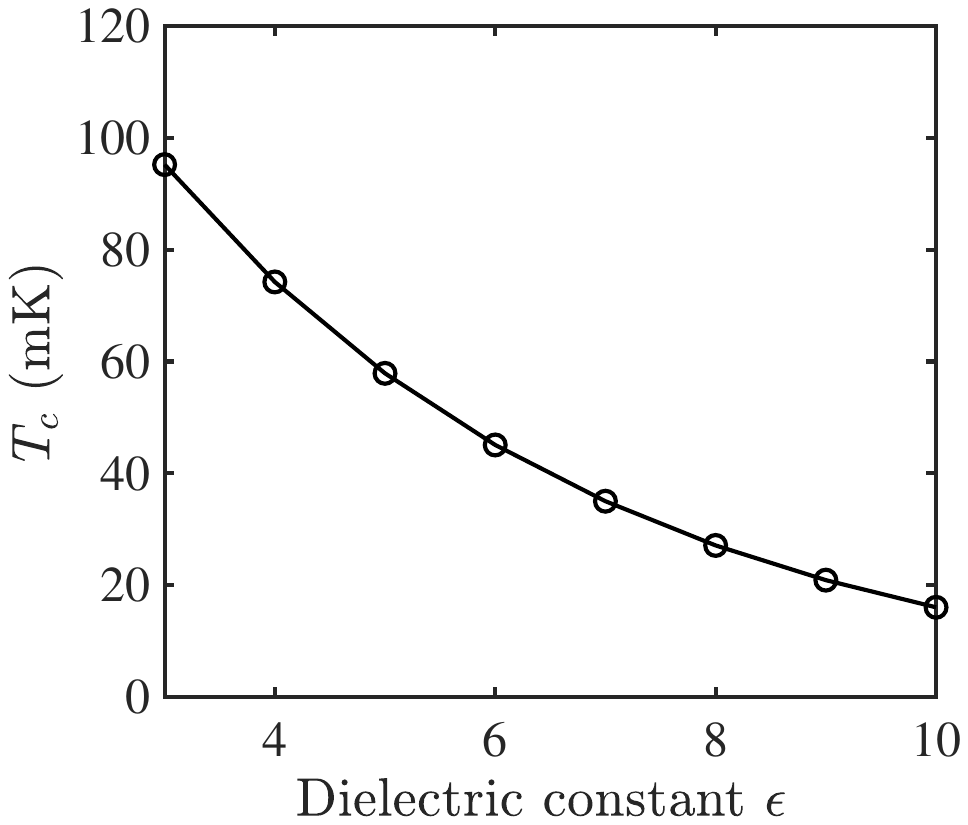}
 \caption{Superconducting critical temperature $T_c$ as a function of dielectric constant $\epsilon$. These results are obtained at the representative phase point indicated by the open arrow in Fig.~3(b) of the main text.}
 \label{fig:figureS10}
\end{figure}

In this study, we adopt a dielectric constant $\epsilon=5$ to approximate the dielectric environment of  tWSe$_2$, which is experimentally realized by encapsulating the system between two hexagonal boron nitride (h-BN) layers \cite{Guo:2024aa}. The dielectric constant of h-BN is approximately 5. To examine the robustness of our results, we have also calculated the superconducting critical temperature as a function of $\epsilon$. As shown in Fig.~\ref{fig:figureS10}, $T_c$ decreases monotonically with increasing $\epsilon$, consistent with the purely electronic mechanism of superconductivity proposed in this work. We have verified that the structure of the superconducting order parameter remains unchanged. 

\section{Off-diagonal correlation function}
\begin{figure}
 \centering
\includegraphics[width=0.55\columnwidth]{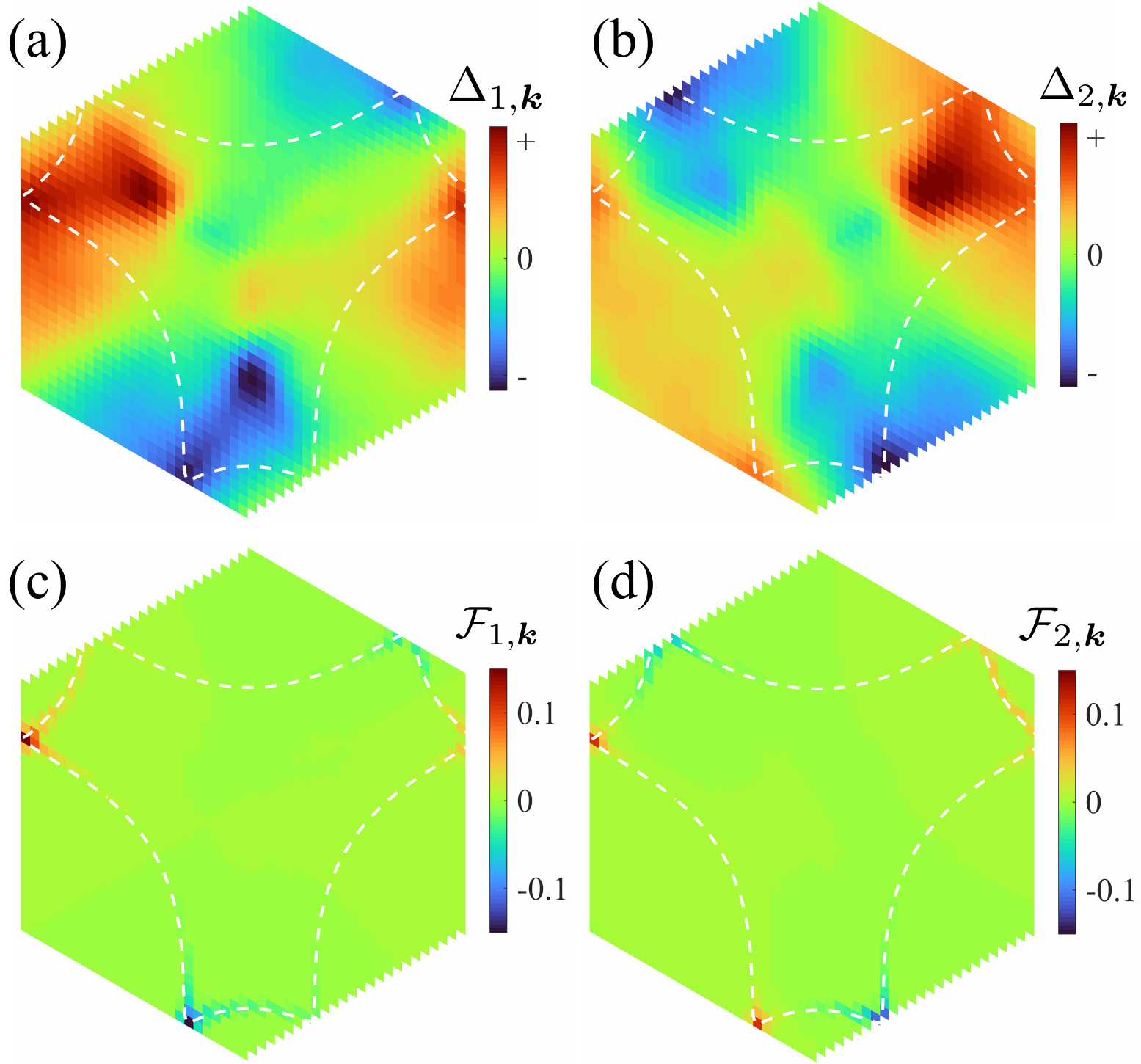}
 \caption{(a),(b) $\bm k$-space structures of the superconducting order parameters $\Delta_{1,\bm k}$ and $\Delta_{2,\bm k}$, reproduced from Figs.~4(a) and (b) of the main text. (c),(d) The corresponding pair correlation function $\mathcal{F}_{1,\bm k}$ and $\mathcal{F}_{2,\bm k}$, calculated at zero temperature. The dashed lines denote the Fermi surface.}
 \label{fig:figureS11}
\end{figure}

As shown in Fig.~4 in the main text (Figs.~4(a) and (b) are reproduced below as Figs.~\ref{fig:figureS11}(a) and (b)), 
the superconducting order parameters exhibit local maxima at the saddle points that contribute to the VHS in the density of states. However, there are also other regions in the MBZ where $\Delta_{\bm k}$ attains comparably large magnitudes. We would like to emphasize that, according to the linearized gap equation (see Eq.~(7) in the main text), the $\bm{k}$-space structure of $\Delta_{\bm k}$ is determined by two key factors: the particle-particle susceptibility, given by $\tanh \left(\beta \epsilon_{\bm k}^K / 2\right) /\left(2 \epsilon_{\bm k}^K\right)$ which is closely related to the VHS, and the effective pairing interaction $V_{\mathrm{eff}}\left(\boldsymbol{k}^{\prime},-\boldsymbol{k}^{\prime};-\boldsymbol{k},\boldsymbol{k}\right)$. While the enhanced density of states around saddle points naturally amplifies $\Delta_{\bm k}$ in thoes region, the specific momentum dependence of the interaction kernel $V_{\mathrm{eff}}$ also plays a critical role in redistributing $\Delta_{\bm k}$ across the MBZ. This interplay may give rise to additional local maxima of $\Delta_{\bm k}$ away from the saddle points.

To clarify this point, we compute the off-diagonal correlation function defined as
\begin{equation}
\mathcal{F}_{\boldsymbol{k}}=\left\langle c_{K,\boldsymbol{k}} c_{K^{\prime},-\boldsymbol{k}}\right\rangle=\frac{\tanh \left(\beta E_{\boldsymbol{k}}^K / 2\right) }{2 E_{\boldsymbol{k}}^K} \Delta_{\boldsymbol{k}},
\end{equation}
which excludes the influence of the interaction kernel. As shown in Figs.~\ref{fig:figureS11}(c) and (d), the resulting $\mathcal{F}_{1,\bm k}$ and $\mathcal{F}_{2,\bm k}$  set exactly on the Fermi surface (indicated by the dashed line), with their maximum values located precisely at the saddle points.
From the gap equation of Eq.~\eqref{eqt27}, it becomes clear that the superconducting order parameter $\Delta_{\bm k}$ arises from the scattering of
$\mathcal{F}_{\bm k'}$, concentrated on the FS, to other regions of the MBZ via the pairing interaction $V_{\mathrm{eff}}\left(\boldsymbol{k}^{\prime},-\boldsymbol{k}^{\prime};-\boldsymbol{k},\boldsymbol{k}\right)$. This scattering mechanism accounts for the broader $\bm{k}$-space distribution of $\Delta_{\bm k}$, including the appearance of local maxima away from the saddle points.
We note that while increasing temperature reduces the magnitudes of $\mathcal{F}_{1,\bm k}$ and $\mathcal{F}_{2,\bm k}$, it has negligible impact on their $\bm{k}$-space structures.

\bibliography{ref}